\documentclass[sigconf]{acmart}

\usepackage{booktabs}
\usepackage{array}
\usepackage{bm}
\usepackage{amsmath}
\usepackage{xcolor}
\usepackage{colortbl}
\usepackage{pifont}
\usepackage{makecell}
\usepackage{multirow}
\usepackage{enumitem}
\usepackage{subcaption}
\usepackage{marvosym}

\newcommand{\yes}{\textcolor{green!70!black}{\ding{51}}}  
\newcommand{\no}{\textcolor{red!70!black}{\ding{55}}}    

\AtBeginDocument{%
  }

\begin{document}


\title{Interactive Recommendation Agent with Active User Commands}

\author{Jiakai Tang\textsuperscript{1}\textsuperscript{$\mathsection$},
Yujie Luo\textsuperscript{3},
Xunke Xi\textsuperscript{3},
Fei Sun\textsuperscript{2},
Xueyang Feng\textsuperscript{1},
Sunhao Dai\textsuperscript{1},
Chao Yi\textsuperscript{3},
Dian Chen\textsuperscript{3},
Zhujin Gao\textsuperscript{3},
Yang Li\textsuperscript{3},
Xu Chen\textsuperscript{1}$^{\textrm{\Letter}}$,
Wen Chen\textsuperscript{3}$^{\textrm{\Letter}}$,
Jian Wu\textsuperscript{3},
Yuning Jiang\textsuperscript{3},
Bo Zheng\textsuperscript{3}$^{\textrm{\Letter}}$
}

\affiliation{%
  \institution{\textsuperscript{1}Gaoling School of Artificial Intelligence, Renmin University of China, Beijing, China}
  \city{\textsuperscript{2}University of Chinese Academy of Sciences, China\\\textsuperscript{3}Alibaba Group, Beijing, China}
  \country{}
}
\email{tangjiakai5704@ruc.edu.cn}
\thanks{$\mathsection$ Work done during internship at Alibaba Group.}
\thanks{${\textrm{\Letter}}$ Corresponding author.}

\renewcommand{\authors}{Jiakai Tang, Yujie Luo, Xunke Xi, Fei Sun, Xueyang Feng, Sunhao Dai, Chao Yi, Dian Chen, Zhujin Gao, Yang Li, Xu Chen, Wen Chen, Jian Wu, Yuning Jiang, Bo Zheng}
\renewcommand{\shortauthors}{Jiakai Tang et al.}
\renewcommand{\shorttitle}{Interactive Recommendation Agent with Active User Commands}

\begin{abstract}
Traditional recommender systems rely on passive feedback mechanisms that limit users to simple choices such as like and dislike. However, these coarse-grained signals fail to capture users' nuanced behavior motivations and intentions. In turn, current systems cannot also distinguish which specific item attributes drive user satisfaction or dissatisfaction, resulting in inaccurate preference modeling. These fundamental limitations create a persistent gap between user intentions and system interpretations, ultimately undermining user satisfaction and harming system effectiveness.

To address these limitations, we introduce the Interactive Recommendation Feed (IRF), a pioneering paradigm that enables natural language commands within mainstream recommendation feeds. Unlike traditional systems that confine users to passive implicit behavioral influence, IRF empowers active explicit control over recommendation policies through real-time linguistic commands.
To support this paradigm, we develop RecBot, a dual-agent architecture where a Parser Agent transforms linguistic expressions into structured preferences and a Planner Agent dynamically orchestrates adaptive tool chains for on-the-fly policy adjustment. To enable practical deployment, we employ simulation-augmented knowledge distillation to achieve efficient performance while maintaining strong reasoning capabilities. Through extensive offline and long-term online experiments, RecBot shows significant improvements in both user satisfaction and business outcomes.

\end{abstract}



\received{20 February 2007}
\received[revised]{12 March 2009}
\received[accepted]{5 June 2009}

\maketitle

\section{Introduction}

\begin{figure}[t]
\centering
\includegraphics[width=\linewidth]{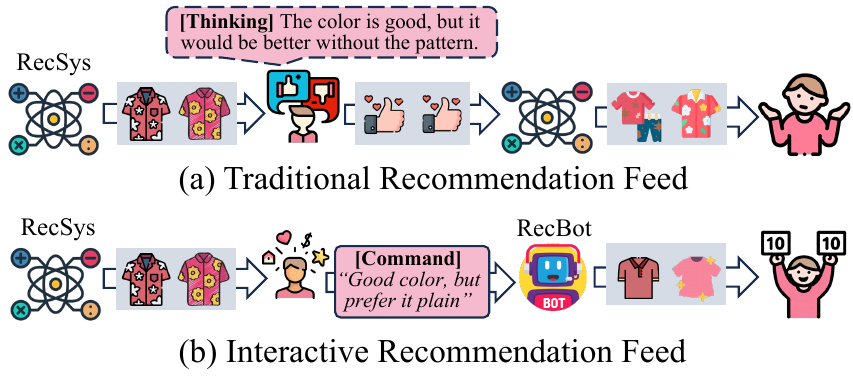}
\vspace{-2em}
\caption{Comparison between traditional and novel interactive recommendation feeds. (a) Traditional systems rely on constrained and implicit feedback signals (\textit{e.g.}, likes/dislikes), making it difficult to accurately infer users' true intentions. (b) Our interactive paradigm enables free-form natural language commands, where RecBot responds and adjusts recommendation policy on-the-fly based on active user commands.}
\vspace{-1em}
\label{fig:intro}
\end{figure}

Modern digital ecosystems are saturated with vast amounts of information, making Recommender Systems (RecSys) indispensable components for managing information overload~\cite{khusro2016recommender,costa2013emotion}, from e-commerce platforms (\textit{e.g.}, Taobao, Amazon)~\cite{yi2025recgpt,smith2017two} to social media feeds (\textit{e.g.}, TikTok, YouTube)~\cite{zhu2024interest,davidson2010youtube}. By filtering and presenting content based on individual user preferences, these systems profoundly shape how users interact with digital content daily.
Existing recommender systems typically operate through a \textit{passive feedback loop}, as illustrated in Fig.~\ref{fig:intro}(a). In this paradigm, users are presented with curated recommendation lists and respond through implicit behavioral signals such as clicks, likes, dislikes, or viewing duration. The recommendation algorithms then analyze these signals to infer user preferences and iteratively refine future recommendations, establishing an iterative implicit feedback loop~\cite{lin2018hybrid,lin2024recrec,wei2024llmrec}.

However, this passive feedback paradigm suffers from fundamental limitations that create a persistent communication gap between user intentions and system interpretations, ultimately undermining recommendation effectiveness. We identify the root causes of this problem from three complementary perspectives: \textit{how users experience interaction constraints, how algorithms interpret ambiguous signals, and how their interplay creates systemic dysfunction.}
First, \textbf{from the user perspective}, current recommendation interfaces restrict users to rigid interaction mechanisms such as click or star ratings, which cannot adequately capture the nuanced reasoning behind user preferences or identify which specific item attributes drive their satisfaction or dissatisfaction. This design limitation forces users to express complex preferences through vague signals that lose critical information about their actual intentions.
Second, \textbf{from the algorithm perspective}, faced with such ambiguous and incomplete user feedback, recommendation algorithms resort to indiscriminate preference attribution, treating all item characteristics as equally responsible for user responses. This approach inevitably leads to inaccurate preference modeling that misrepresent user intentions and contribute to filter bubbles that narrow content diversity.
Third, \textbf{from the ecosystem perspective}, these user expression constraints and algorithmic interpretation flaws interact to create cascading dysfunction beyond their respective impacts. When systems misinterpret signals and users respond to irrelevant recommendations, this creates a communication deadlock where neither party can effectively convey or understand the other's intentions, necessitating fundamental paradigm change rather than incremental improvements.

To overcome these limitations, we first propose a novel \textbf{Interactive Recommendation Feed (IRF)} product paradigm from a user-centric design philosophy, as illustrated in Fig.~\ref{fig:intro}(b). 
Unlike traditional interactive recommendation that confine users to passive consumption~\cite{shi2024large,gao2023alleviating,zhou2020interactive}, IRF introduces an active voice channel enabling users to directly interact with the recommender systems through free-form natural language commands. This transforms the conventional one-way delivery into bidirectional interaction where users can explicitly articulate their requirements and constraints to guide the system's policy adjustment in real-time. 
Moreover, IRF also differs intrinsically from Conversational Recommender Systems (CRS), which require independent dialogue windows for guided questioning. Instead, IRF seamlessly integrates into mainstream recommendation feeds  (e.g., Youtube, Taobao) as a lightweight interface, allowing direct commands without interrupting natural browsing flow.
By empowering users with direct algorithmic control, IRF essentially rethinks the human-system relationship, enabling users to proactively shape their content consumption experience rather than being passive recipients of algorithmic decisions.

Furthermore, accurately responding to textual user commands requires bridging the gap between open-ended user expressions and actionable recommendation strategies, yet existing interactive methods lack the natural language understanding and reasoning capabilities necessary to parse free-form commands and translate them into command-aware item scoring adjustments. To tackle this problem, we develop \textbf{\textit{RecBot}}, a large language model-powered multi-agent framework for interactive recommender systems.
The framework comprises two core components: First, the \textbf{User-Intent-Understanding Agent (\textit{Parser})} parses natural language descriptions into structured recommendation specifications, extracting both positive and negative user intents while filtering irrelevant information. Second, the \textbf{Planning Agent (\textit{Planner})} then receives these structured preferences and orchestrates customized tool invocation chains, transforming domain-specific requirements into concrete actions that modify item relevance scores. This tool-chaining approach enables the Planner to handle sophisticated user intentions (\textit{e.g.}, positive interest, negative feedback, and interest drift), facilitating on-the-fly policy adaptation that directly influences next recommendation feed. 
To better support production deployment, RecBot incorporates dynamic memory consolidation for efficient multi-turn interactions and employs knowledge distillation to transfer understanding and reasoning capabilities from powerful teacher models to lightweight student models, achieving scalable and cost-effective deployment.

To comprehensively evaluate the effectiveness of our proposed Interactive Recommendation Feed paradigm and the RecBot framework, we conduct extensive empirical validation through \textbf{both offline and online experiments}. Our offline evaluation spans three real-world recommendation datasets (Amazon, MovieLens, and Taobao), where RecBot demonstrates superior performance against numerous state-of-the-art baseline methods across multiple evaluation metrics. 
RecBot has been \textbf{fully deployed} in a production environment within a large-scale e-commerce platform's homepage, with a three-month online A/B testing evaluation. The \textbf{long-term online} experimental results reveal substantial improvements across multiple key metrics, including enhanced user engagement (\textit{e.g.}, \textbf{0.71\%} reduction in Negative Feedback Frequency (NFF) and \textbf{1.44\%} increase in Click Item Category Diversity (CICD)) and significant business revenue gains (\textit{e.g.}, \textbf{1.28\%} increase in Add-to-Cart (ATC) and \textbf{1.40\%} increase in Gross Merchandise Volume (GMV)). 
Our contributions are summarized as follows:
\begin{itemize}[leftmargin=*]
  \item We introduce the novel Interactive Recommendation Feed (IRF) product paradigm that \textbf{breaks the silence of recommender systems} by allowing users to directly communicate with the system through natural language commands, achieving user-centric controllable recommendation.
  \item We propose RecBot, a multi-agent framework comprising the \textit{Parser} and \textit{Planner} agents that achieve precise intent parsing and on-the-fly command-aware recommendation policy adaptation through automated tool chaining, effectively bridging the gap between free-form user commands and actionable adjustments.
  \item Extensive offline and online experiments are conducted to comprehensively validate RecBot's effectiveness. Notably, our online deployment demonstrates significant improvements in user engagement, content diversity, and business benefits.
\end{itemize}

\section{Problem Setup and Formulation}
In this section, we introduce the \textbf{\textit{Interactive Recommendation Feed (IRF)}} paradigm, which formulates the recommendation process as a sequential decision-making problem where the system iteratively refines recommendations based on user linguistic commands and behavioral signals. Let \( I = \{ i_1, i_2, \dots, i_n \} \) denote the candidate item pool, and \( u \) represent a target user. At each interaction round \( t \in \{ 1, 2, \dots, T \} \), the system presents a recommendation feed \( R_t = \{ r^{(1)}_t, r^{(2)}_t, \dots, r^{(K)}_t \} \subset I \) of \( K \) items to the user. Upon receiving \( R_t \), the user provides linguistic feedback \( c_t \). This feedback captures the user's assessment of current recommendations, preferences, or any specific constraints or requirements. 

The system maintains a \textbf{dual-component user model}: \textit{Explicit Preferences} \( P_t \) and \textit{Implicit Preferences} \( H_t \). Explicit preferences represent structured knowledge extracted from user linguistic feedback through natural language understanding, formally defined as \( P_t = \phi(c_1, c_2, \dots, c_t) \), where \( \phi \) maps a sequence of linguistic feedbacks to structured preference representations. Implicit preferences are captured by the user's historical interaction sequence \( H_t = \{ i_1, i_2, \dots, i_m \} \), reflecting long-term behavioral patterns.

At round \( t \), the system state is defined as \( S_t = \{ R_t, c_t, P_t, H_t \} \). The system's action involves selecting the next recommendation feed: \( R_{t+1} = \pi(S_t) \). The IRF process operates iteratively: beginning with initial feed \( R_0 \), the system collects feedback \( c_t \), updates explicit preferences \( P_t \) while incorporating historical behavior \( H_t \), constructs state \( S_t \), and generates the next feed $R_{t+1}$. This continues until user satisfaction is achieved $ \mathrm{satisfy}(u, R_{t+1}) = 1 $ or the maximum interaction limit is reached $ t \geq T_{\max} $, where $ T_{\max} $ represents the threshold for user engagement fatigue beyond which users typically disengage from the system.

\begin{figure}[t]
\centering
\includegraphics[width=\linewidth]{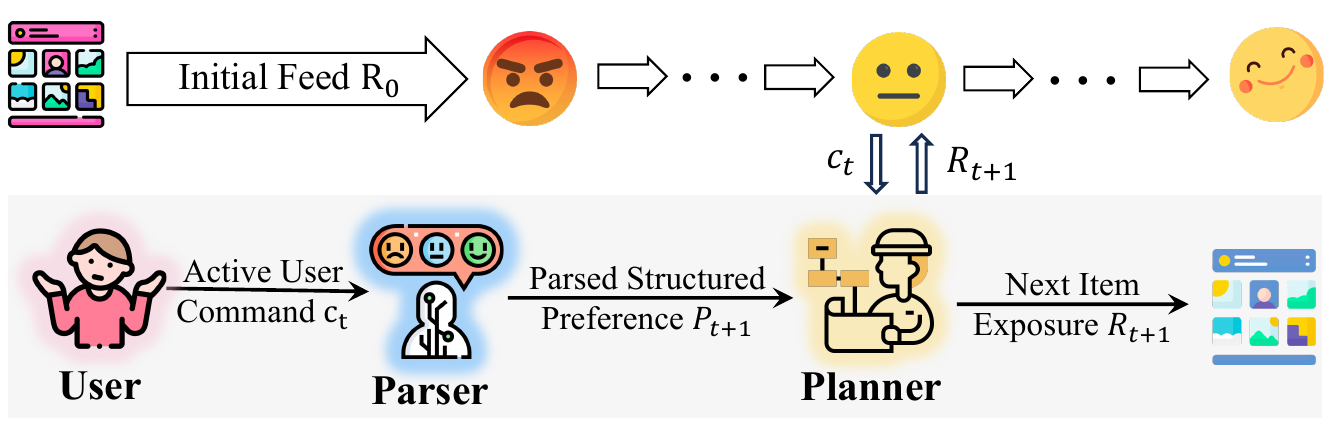}
\vspace{-2em}
\caption{Overview of the RecBot framework for interactive recommendation. The framework comprises a Parser Agent that transforms user natural language command $c_t$ into structured preferences $P_{t+1}$, and a Planner Agent that orchestrates tool chains to dynamically adjust recommendation policies and generate the next feed $R_{t+1}$.}
\label{fig:framework}
\vspace{-2em}
\end{figure}

\section{Methodology}
This section first provides an overview of our proposed method, then elaborates each core component, and finally presents our multi-agent optimization approach for lightweight deployment.

\subsection{Overall Architecture}
To enable precise responsiveness to user-initiated active commands, we introduce \textbf{RecBot}, a multi-agent framework that systematically decomposes multi-turn user commands, dynamically adjusts recommendation strategies, and modifies subsequent recommendation feeds to progressively satisfy user requirements. The RecBot framework comprises two core intelligent agents coupled with a comprehensive and extensible tool suite, facilitating the transformation from free-form natural language commands to actionable item scoring computations.

As illustrated in Fig.~\ref{fig:framework}, our approach operates through a sequential two-stage process. First, the \textbf{User-Intent-Understanding Agent (\textit{Parser})} transforms user open-domain textual instructions into structured domain-specific recommendation language, parsing both positive and negative user intents while employing a dynamic memory consolidation prompting strategy to maintain efficient multi-turn dialogue state management. Subsequently, the \textbf{Planning Agent (Planner)} receives the structured recommendation specifications and orchestrates appropriate tool chain invocations, ultimately curating and presenting the top-$K$ items to users. This systematic decomposition enables RecBot to bridge the semantic gap between diverse user expressions and concrete recommendation adjustments, facilitating real-time policy adaptation through principled tool orchestration.

\subsection{Parser}
Real-world user utterances exhibit considerable variability and diversity in both style and semantic content~\cite{tang2025kapa,zhang2024towards}, often containing redundant noise such as verbose complaints, irrelevant contextual information. The key challenge lies in transforming free-form user commands into structured recommendation instructions, which is essential for accurately capturing users’ proactive intentions.
Inspired by the robust contextual reasoning capabilities of large language models~\cite{sui2025stop,li2025system,patil2025advancing,xu2025towards}, we design a \textbf{User Intent Understanding Agent (\textit{Parser})} that bridges the semantic gap between diverse user expressions and actionable recommendation specifications. The \textit{Parser} exploits the inherent ability of large language models to handle linguistic variability and extract meaningful signals from noisy textual input, enabling precise intent understanding.

\subsubsection{\textbf{Structured Command Parsing}}
Given the previous recommendation feed $R_t$ presented to the user, user linguistic feedback $c_t$ in response to $R_t$, and historical preference memory $P_t$, the \textit{Parser} performs the transformation:
\begin{equation}
\mathcal{P} : (R_t, c_t, P_t) \rightarrow P_{t+1},
\end{equation}
where $\mathcal{P}$ denotes the parsing function that maps the input tuple to updated preference specifications $P_{t+1}$. The updated preference specification $P_{t+1}$ decomposes into two orthogonal dimensions based on user sentiment orientation:
\begin{equation}
P_{t+1} = \{ P_{t+1}^+, P_{t+1}^- \}.
\end{equation}
where $P_{t+1}^+$ represents \textbf{positive preferences} (item attributes the user shows interest in) and $P_{t+1}^-$ represents \textbf{negative preferences} (item attributes the user wants to exclude). This bidirectional decomposition enables better modeling of user intentions by capturing both attraction and aversion signals from natural language feedback.
Furthermore, each preference dimension further categorizes into constraint types based on enforcement strictness:
\begin{equation}
P_{t+1}^+ = \{ C_{t+1}^{+,\mathrm{hard}}, C_{t+1}^{+,\mathrm{soft}} \}, \quad 
P_{t+1}^- = \{ C_{t+1}^{-,\mathrm{hard}}, C_{t+1}^{-,\mathrm{soft}} \},
\end{equation}
where $C^{\mathrm{hard}}$ represents \textbf{strict rule-based constraints} that can be deterministically verified through direct attribute matching (\emph{e.g.}, price thresholds: ``under \$50''), and $C^{\mathrm{soft}}$ denotes \textbf{flexible semantic-based inclinations} that require nuanced understanding of user's subjective interests (\emph{e.g.}, genre preference: ``prefer romantic movies'').

\paragraph{\textbf{Remark}}
Unlike existing interactive recommendation methods that predominantly focus on positive user requirements~\cite{wu2023goal,wu2024personalised,huang2025towards}, \textit{Parser} explicitly models both positive and negative preference signals. This bidirectional approach is motivated by empirical observations in production environments, where negative feedback accounts for the majority of user interactions\footnote{In-house platform deployment analysis reveals negative feedback accounts for approximately 57\% of user commands versus 43\% positive feedback.}, necessitating comprehensive preference representation that encompasses the full spectrum of user intentional expressions.

\subsubsection{\textbf{Dynamic Memory Consolidation}}
Multi-turn interactions are crucial for interactive recommender systems as they allow users to iteratively refine their preferences and guide the system toward more in-depth understanding of their evolving interests.
However, multi-turn interactions introduce extra computational and semantic challenges in maintaining consistent user preference understanding across extended dialogue sequences. Naive approaches that accumulate extensive conversation histories lead to computational overhead and semantic drift, while oversimplified state management fails to capture evolving user intentions accurately.
To this end, we design a \textbf{Dynamic Memory Consolidation Strategy} that maintains preference coherence across interaction rounds while ensuring computational efficiency through principled state management rather than exhaustive history retention.

In specific, the consolidation mechanism operates through three \textbf{strategic update principles} that guide the Parser's preference synthesis process:

\textbf{(1) Preservation Principle}: When current feedback $c_t$ indicates satisfaction with existing recommendations or provides neutral commentary, historical preferences $P_t$ are preserved unchanged to maintain preference stability:
$P_{t+1} = P_t.$

\textbf{(2) Integration Principle}: When feedback introduces compatible new preferences that complement existing memory, these preferences are integrated through structured merging operations:
\[
P_{t+1}^+ = P_t^+ \oplus \text{Extract}^+(c_t), 
\qquad
P_{t+1}^- = P_t^- \oplus \text{Extract}^-(c_t),
\]
where $\oplus$ denotes the structured integration operator that handles both hard constraints and soft preferences separately, and $\text{Extract}^\pm(c_t)$ extracts positive and negative preference signals from current user command $c_t$ respectively.

\textbf{(3) Resolution Principle}: When feedback explicitly contradicts established preferences, the Parser employs linguistic cue analysis to identify change indicators (e.g., ``instead of,'' ``no longer interested in,'' ``different from before'') and performs targeted preference updates while preserving non-conflicting knowledge:
\[
P_{t+1} = \text{Resolve}(P_t, c_t, R_t).
\]
The consolidation process is formalized through a \textbf{context-aware decision function} that dynamically selects the appropriate update strategy based on the interplay between historical preferences $P_t$, current feedback $c_t$, and recommendation context $R_t$:
\[
P_{t+1} =
\begin{cases}
P_t & \text{if } \phi_{\text{sat}}(c_t, R_t), \\[4pt]
P_t \oplus \text{Extract}(c_t) & \text{if } \phi_{\text{com}}(P_t, c_t, R_t), \\[4pt]
\text{Resolve}(P_t, c_t, R_t) & \text{if } \phi_{\text{con}}(P_t, c_t, R_t),
\end{cases}
\]
where $\phi_{\text{sat}}, \phi_{\text{com}},$ and $\phi_{\text{con}}$ represent context-aware analysis functions that respectively detect user satisfaction states, preference compatibility, and preference conflicts.

\paragraph{\textbf{Remark}}
The proposed dynamic memory consolidation approach concentrates large language model reasoning on synthesizing compressed historical preferences with current feedback, avoiding repetitive processing of extensive multi-turn conversation histories while maintaining information fidelity across extended interaction rounds with bounded computational complexity.

\begin{figure}
\centering
\includegraphics[width=\linewidth]{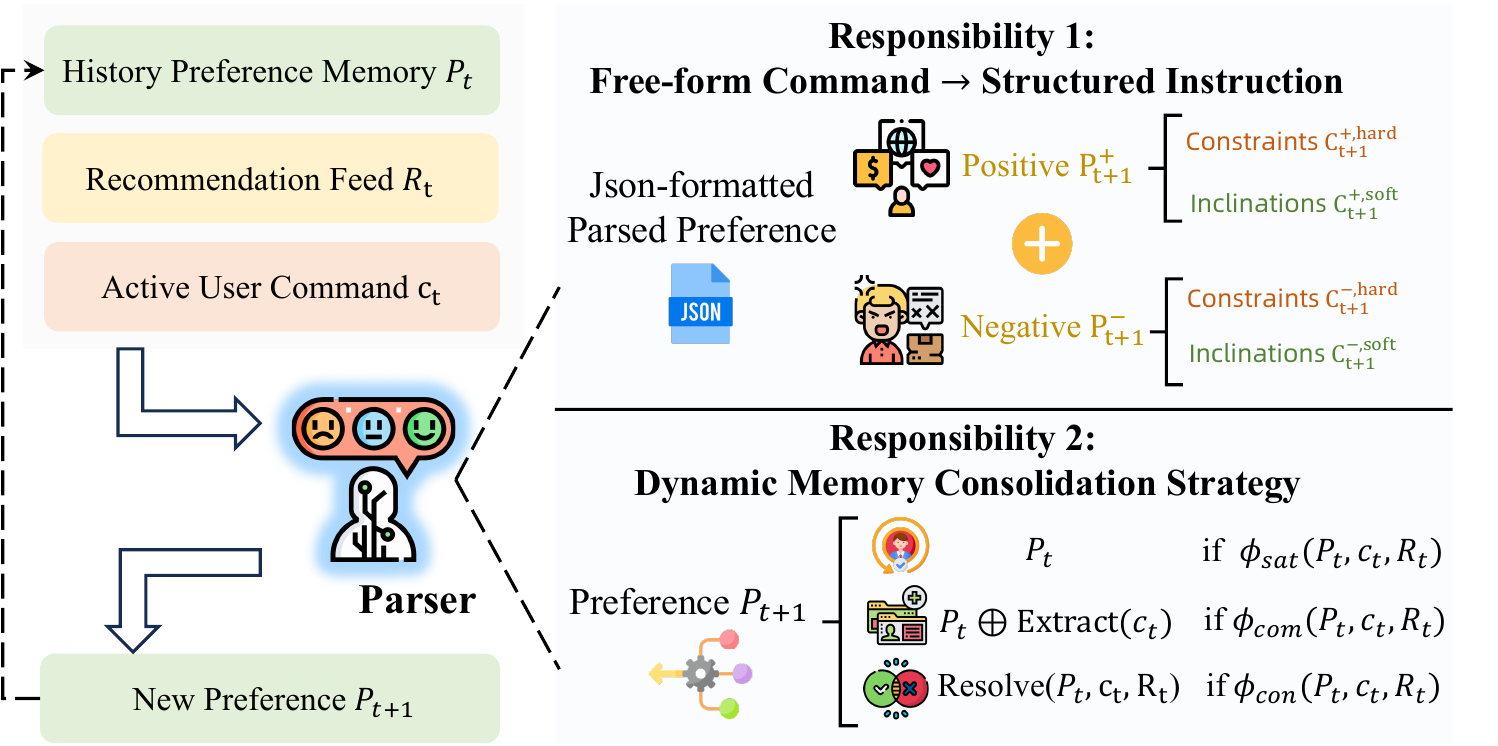}
\vspace{-2em}
\caption{Illustration of the \textit{Parser} for user intent understanding. The \textit{Parser} integrates history preference memory $P_t$, current recommendation feed $R_t$, and active user command $c_t$ to generate new preference representation $P_{t+1}$ through structured parsing and dynamic memory consolidation.}
\label{fig:parser}
\vspace{-1em}
\end{figure}

\subsection{Planner}
Upon receiving the structured recommendation specifications $P_{t+1}$ from the \textit{Parser}, the next task is to translate these explicit user preferences into concrete recommendation policy adjustments. To address this, we introduce the \textbf{Planning Agent (\textit{Planner})}, which orchestrates adaptive tool chain invocations to dynamically modify item scoring mechanisms based on user commands.

The \textit{Planner} operates through principled reasoning over a comprehensive \textbf{recommendation domain toolset}, enabling flexible adaptation to diverse user requirements. Given structured preferences $P_{t+1} = \{P_{t+1}^+, P_{t+1}^-\}$ and user implicit preferences $H_t$, the \textit{Planner} performs the mapping:
\begin{equation}
\mathcal{A}: (P_{t+1}, H_t, I) \rightarrow S_{t+1},
\end{equation}
where $\mathcal{A}$ denotes the planning function that transforms preference specifications and candidate item pool $I$ into updated item scoring $S_{t+1} = \{s_1, s_2, \ldots, s_n\}$, enabling top-$K$ item selection for the subsequent recommendation feed $R_{t+1}$.

\subsubsection{\textbf{Recommendation Domain Toolset}}

The \textit{Planner} uses a modular and extensible toolset, each designed to handle specific aspects of user preference satisfaction through targeted item scoring.
The core toolset comprises four complementary components: (i) \textbf{\textit{Filter}} for hard constraint enforcement, (ii) \textbf{\textit{Matcher}} for positive preference alignment, (iii) \textbf{\textit{Attenuator}} for negative feedback incorporation, and (iv) \textbf{\textit{Aggregator}} for comprehensive score integration.
In the following, we detail the specific functionality of each tool.

\paragraph{\textbf{Filter Tool}}
The \textit{Filter} enforces hard constraint satisfaction by both selecting items that meet positive requirements and eliminating items that violate negative constraints. Given hard constraints $C_{t+1}^{+,\mathrm{hard}}$ and $C_{t+1}^{-,\mathrm{hard}}$ from $P_{t+1}$, the filter adopts binary selection over the candidate item pool $I$ to produce a refined subset $I'$:
\begin{equation}
I' = \{i \in I : \mathcal{C}^+(i, C_{t+1}^{+,\mathrm{hard}}) = 1 \land \mathcal{C}^-(i, C_{t+1}^{-,\mathrm{hard}}) = 0\},
\end{equation}
where $\mathcal{C}^+(i, C_{t+1}^{+,\mathrm{hard}})$ evaluates whether item $i$ satisfies all positive hard constraints, and $\mathcal{C}^-(i, C_{t+1}^{-,\mathrm{hard}})$ determines whether item $i$ violates any negative hard constraints. Items excluded from $I'$ are effectively assigned a score of $-\infty$, ensuring they are not recommended in the subsequent feed.

This filtering mechanism enables efficient candidate space pruning while enforcing strict user requirements. For instance, when users specify temporal constraints (\textit{e.g.}, ``movies released before 2020'') or budget limitations (\textit{e.g.}, ``no luxury products above \$500''), the \textit{Filter} precisely narrows the search space according to these explicit boundaries. The resulting filtered candidate set $I'$ serves as the refined input for downstream scoring tools, namely the \textit{Matcher} and \textit{Attenuator}, ensuring computational efficiency while maintaining strict adherence to user-specified requirements.

\paragraph{\textbf{Matcher Tool}}
The \textit{Matcher} tool computes positive relevance scores based on user positive intentions through a two-path architecture that combines semantic understanding with personalized collaborative knowledge.

\textbf{(1) \textit{Semantic Relevance Path:}} Positive intentions $P_{t+1}^+$ are first transformed into structured natural language descriptions following the format:
``$\mathrm{attribute}_1\mathrm{:} [\mathrm{value}, \ldots], \mathrm{attribute}_2\mathrm{:} [\mathrm{value}, \ldots].$''
The semantic component leverages pre-trained embedding model (\textit{e.g.}, BGE~\cite{xiao2024c}, Sentence-BERT~\cite{reimers2019sentence}) to compute semantic affinity between item descriptions and positive intent representations:
\begin{equation}
s_{\mathrm{sem}}(i, P_{t+1}^+) = \mathrm{sim}(\mathbf{e}_{\mathrm{item}}(i), \mathbf{e}_{\mathrm{intent}}(P_{t+1}^+))
\end{equation}
where $\mathbf{e}_{\mathrm{item}}(i)$ and $\mathbf{e}_{\mathrm{intent}}(P_{t+1}^+)$ denote the embedding representations of item $i$'s description and formatted positive intentions, respectively, and $\mathrm{sim}(\cdot, \cdot)$ indicates cosine similarity.

\textbf{(2) \textit{Active-Intent-Aware Collaborative Path:}} To incorporate personalized collaborative information, we design an \textbf{Active-Intent-Aware (AIA)} sequential recommender that treats user positive feedback as queries to extract intent-relevant patterns from historical multimodal item representations.
Specifically, the AIA model first processes multimodal item features through dedicated encoders $f_m(\cdot)$, where $m$ denotes modality type (\textit{e.g.}, text and image), to obtain modality-specific item representations $\mathbf{h}_m^{i} \in \mathbb{R}^d$.
These are then fused into unified multimodal representations:
\begin{equation*}
\mathbf{h}_{\mathrm{fused}}^{i} = \mathrm{Linear}([\mathbf{h}_{\mathrm{text}}^{i} \oplus \mathbf{h}_{\mathrm{image}}^{i} \oplus \ldots]),
\end{equation*}
where $\oplus$ denotes concatenation and Linear($\cdot$) represents a linear network that performs dimension reduction from the concatenated multimodal representations to unified $d$-dimensional representations.
Similarly, user positive intent $P_{t+1}^+$ is encoded into the textual modality representation $\mathbf{h}_{\mathrm{intent}} \in \mathbb{R}^d$.

Subsequently, modality-specific sequential encoders extract temporal patterns from user historical interactions $H_t$:
\begin{equation*}
\mathbf{H}_{\mathrm{fused}} = \mathrm{SeqEnc}([\mathbf{h}^{i_1}_{\mathrm{fused}}, \mathbf{h}^{i_2}_{\mathrm{fused}}, \ldots, \mathbf{h}^{i_{|H_t|}}_{\mathrm{fused}}]),
\end{equation*}
where $\mathrm{SeqEnc}$ employs Multi-Head Self-Attention (MHSA) mechanisms.
To further capture intent-aware collaborative patterns, we utilize Multi-Head Cross-Attention (MHCA) between transformed user intent and fused sequence representations:
\begin{equation*}
s_{\mathrm{aia}}(i, P_{t+1}^+, H_t) = \mathrm{MHCA}(\mathbf{h}_{\mathrm{intent}}, \mathbf{H}_{\mathrm{fused}}, \mathbf{H}_{\mathrm{fused}}) \cdot \mathbf{h}_{\mathrm{fused}}^{i},
\end{equation*}
where $\mathrm{MHCA}(\mathbf{q}, \mathbf{K}, \mathbf{V})$ computes attention-weighted representations using query $\mathbf{q}$, keys $\mathbf{K}$, and values $\mathbf{V}$.

The final positive relevance score integrates both semantic and collaborative components through weighted summation:
\begin{equation}
s_{\mathrm{match}}(i) = \alpha \cdot s_{\mathrm{sem}}(i, P_{t+1}^+) + (1-\alpha) \cdot \mathcal{S}_{\mathrm{aia}}(i, P_{t+1}^+, H_t),
\end{equation}
where $\alpha \in [0,1]$ balances semantic and collaborative contributions.

\paragraph{\textbf{Attenuator Tool}}
The \textit{Attenuator} computes negative relevance penalties based on user negative intentions. Similar to the semantic path in \textit{Matcher}, negative intentions $P_{t+1}^-$ are converted to structured descriptions and semantic similarity scores are computed:
\begin{equation}
s_{\mathrm{atten}}(i) = -\beta \cdot \mathrm{sim}(\mathbf{e}_{\mathrm{item}}(i), \mathbf{e}_{\mathrm{intent}}(P_{t+1}^-)),
\end{equation}
where $\beta > 0$ controls the attenuation strength, and the negative sign ensures that items semantically similar to negative intentions receive penalty scores, thereby reducing their likelihood of being recommended in the next feed.

\paragraph{\textbf{Aggregator Tool}}
The \textit{Aggregator} tool synthesizes scores from all preceding tools to produce final item rankings:
\begin{equation}
s_{\mathrm{final}}(i) = s_{\mathrm{match}}(i) + s_{\mathrm{atten}}(i),
\end{equation}
Finally, the top-$K$ items based on final scores $s_{\mathrm{final}}(i)$ constitute the next recommendation feed $R_{t+1}$.

\paragraph{\textbf{Remark}}
The modular toolset design enables the \textit{Planner} to flexibly orchestrate semantic understanding, collaborative filtering, and constraint satisfaction through standardized tool interfaces. This design philosophy aligns with emerging paradigms such as \textit{Model Context Protocol (MCP)}~\cite{hou2025model,ray2025survey}, which advocates for structured tool integration in AI systems. For example, a \textit{\textbf{Searcher}} tool for online information retrieval—when user feedback contains trending topics or seasonal preferences—can be seamlessly integrated following MCP principles without requiring fundamental framework modifications. This comprehensive foundation facilitates responsive recommendation policy adaptation that effectively bridges explicit user commands with personalized implicit preferences.

\begin{figure}
\centering
\includegraphics[width=\linewidth]{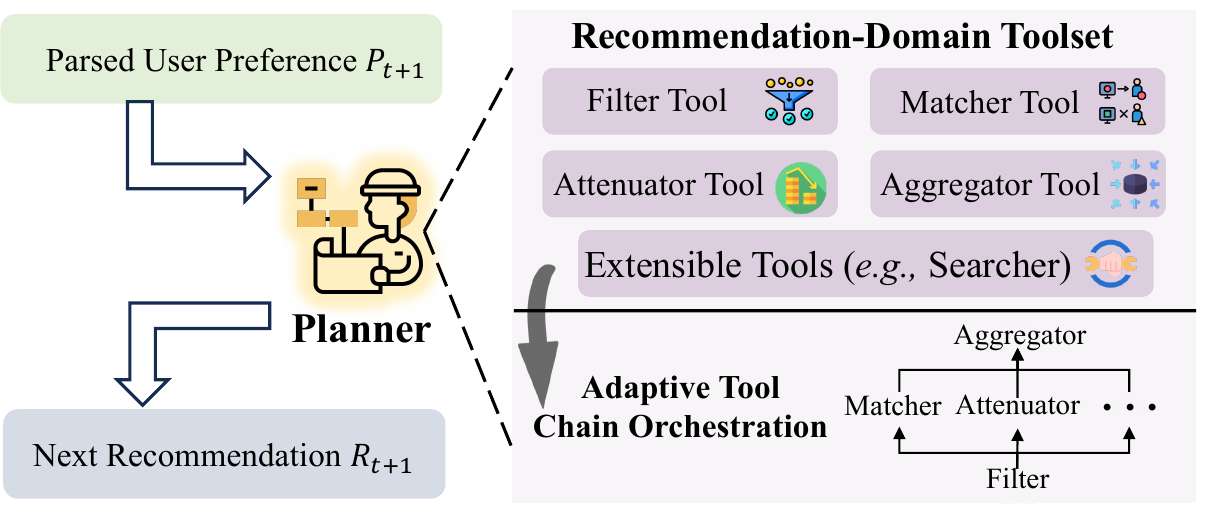}
\vspace{-2em}
\caption{Illustration of the \textit{Planner} for on-the-fly recommendation policy adaptation. The \textit{Planner} dynamically constructs optimal tool invocation sequences based on parsed user preferences $P_{t+1}$ to compute updated item scores $s_{\mathrm{final}}$ for next recommendation feed $R_{t+1}$.}
\label{fig:planner}
\vspace{-1em}
\end{figure}

\subsubsection{\textbf{Adaptive Tool Chain Orchestration}}
The \textit{Planner} employs context-aware reasoning to dynamically construct optimal tool invocation sequences based on the parsed user preferences \( P_{t+1} \). This orchestration operates through two coordinated phases: strategic tool selection and execution coordination.
\begin{itemize}[leftmargin=*]
  \item \textbf{Strategic Tool Selection}
The \textit{Planner} systematically analyzes \( P_{t+1} \) to determine necessary tool activations and execution strategies. Hard constraints \( (C^{+,\mathrm{hard}}_{t+1}, C^{-,\mathrm{hard}}_{t+1}) \) trigger prioritized \textit{Filter} activation to establish refined candidate space \( I' \subset I \). Positive intentions \( (C^{+,\mathrm{soft}}_{t+1}) \) activate the \textit{Matcher} for relevance computation, while negative feedback \( (C^{-,\mathrm{soft}}_{t+1}) \) necessitates \textit{Attenuator} deployment for penalty scoring. The \textit{Planner} adaptively scales tool chain complexity: deploying selective tool subsets for focused feedback to minimize computational overhead, or activating full sequences (\textit{Filter} $\rightarrow$ \textit{Matcher} \& \textit{Attenuator} $\rightarrow$ \textit{Aggregator}) for multi-dimensional user requirements containing both preferences and constraints.
  \item \textbf{Execution Coordination}
The \textit{Planner} manages seamless information flow between activated tools through structured dependency management. The filtered candidate set \( I' \) serves as input constraint for both \textit{Matcher} and \textit{Attenuator}, ensuring all scoring computations operate within the valid item space defined by hard constraints. When applicable, \textit{Matcher} and \textit{Attenuator} execute in parallel to optimize computational throughput, with the \textit{Aggregator} synthesizing multi-tool outputs through weighted combination strategies to produce unified item scores reflecting comprehensive user reference satisfaction.
\end{itemize}

\subsection{Multi-Agent Optimization}\label{sec:optimization}
The \textit{Parser} and \textit{Planner} agents face highly complex personalized reasoning and decision-making tasks involving ambiguous user command parsing, multi-turn preference state maintenance, and adaptive tool chain orchestration. Our empirical experiments reveal that closed-source large language models (\textit{e.g.}, GPT-4.1~\cite{achiam2023gpt}) significantly outperform open-source alternatives in these critical tasks. However, these LLMs face deployment constraints (\textit{e.g.}, prohibitive inference costs, data privacy concerns, and limited controllability). To achieve lightweight online deployment while reconciling this performance-deployment tension, we introduce a \textbf{Multi-Agent Optimization} framework that facilitates knowledge transfer from powerful closed-source teacher models to cost-effective open-source student models through simulation-augmented knowledge distillation method.

\subsubsection{\textbf{Simulation-Augmented Knowledge Distillation Framework}}
Our optimization approach leverages the versatile role-playing capabilities~\cite{gu2024survey,zheng2023judging,pan2024human} of large language models to construct a \textbf{simulation-based training environment} where synthetic user agents interact with teacher RecBot systems across multiple interaction rounds. This paradigm enables diverse data generation that captures the varied user-system interactions while maintaining authentic dialogue patterns and preference dynamics.

We establish a simulation framework where a \textbf{User Simulation Agent} $\mathcal{U}_{\mathrm{sim}}$ and a \textbf{Teacher RecBot} $\mathcal{R}_{\mathrm{teacher}}$ engage in multi-turn interactive processes. Given a target item anchor $i_{\mathrm{target}} \in I$ representing the user's latent objective and an initial recommendation feed $R_0$, the simulation proceeds iteratively: the User Simulation Agent generates contextually appropriate linguistic feedback following natural user expression patterns:
\begin{equation}
  c_t^{\mathrm{sim}} = \mathcal{U}_{\mathrm{sim}}(R_t, i_{\mathrm{target}}, \mathcal{G}_{\mathrm{persona}}),
\end{equation}
where $\mathcal{G}_{\mathrm{persona}}$ represents persona specifications that define user behavioral characteristics, including preference expression styles, constraint specification patterns, and interaction tendencies. 

Then, $\mathcal{R}_{\mathrm{teacher}}$ processes this feedback through both \textit{Parser} and \textit{Planner} components, generating structured preferences and corresponding tool chains for recommendation adaptation. This process continues until the user agent reaches satisfaction or meets termination criteria (\textit{e.g.}, maximum interaction rounds), yielding multi-turn interaction trajectories that capture varied user scenarios, preference evolution patterns, and system response strategies across different contexts.

\subsubsection{\textbf{Training Data Collection}}
From the generated simulation trajectories, we systematically extract training samples for both \textit{Parser} and \textit{Planner} agents. For the \textbf{\textit{Parser}}, each training sample captures the complete preference parsing transformation: input tuple $(R_t, c_t^{\mathrm{sim}}, P_t)$ and corresponding structured output $P_{t+1}^{\mathrm{sim}}$. For the \textbf{\textit{Planner}}, training instances demonstrate adaptive tool orchestration: structured preference specifications $P_{t+1}^{\mathrm{sim}}$ paired with optimal tool invocation sequences $\mathcal{T}_t^{\mathrm{sim}}$ determined by the teacher model. The training datasets are formally defined as:
\[
\mathcal{D}_{\mathrm{Parser}} = \{((R_t, c_t^{\mathrm{sim}}, P_t), P_{t+1}^{\mathrm{sim}})\}, \quad
\mathcal{D}_{\mathrm{Planner}} = \{((P_{t+1}^{\mathrm{sim}}, \Omega), \mathcal{T}_t^{\mathrm{sim}})\},
\]
where $\Omega$ represents available recommendation tool descriptions. 
Here, to achieve efficient online inference, we implement the \textit{Parser} as an end-to-end function that jointly performs user command interpretation and dynamic memory consolidation in one pass.

\subsubsection{\textbf{Optimization Objective}}
We optimize both agents jointly through Supervised Fine-Tuning (SFT) on a unified language model backbone using a mixed training dataset that combines both Parser and Planner training samples.
Specifically, the unified training dataset is constructed by merging both agent-specific samples: $\mathcal{D}_{\mathrm{Mixed}} = \mathcal{D}_{\mathrm{Parser}} \cup \mathcal{D}_{\mathrm{Planner}}$.
The optimization objective follows standard Next-Token Prediction (NTP) learning paradigm, that is, minimizing the Negative Log-Likelihood (NLL) of target sequences:
\begin{equation}
\mathcal{L}(\theta) = \sum_{(x,y) \in \mathcal{D}_{\mathrm{Mixed}}} \sum_{j=1}^{|y|} -\log P_\theta(y_j | x, y_{<j})
\end{equation}
where $\theta$ represents the shared model parameters, and $y$ denotes the target output sequence (structured preferences for Parser tasks or tool chains for Planner tasks).

This unified training paradigm leverages the model's capacity to handle diverse reasoning patterns through appropriate input formatting and task-specific prompting, enabling cost-effective deployment while maintaining functional modularity. In the future, we will explore independent agent training method to achieve enhanced task-specific performance and more advanced multi-agent learning approaches (\textit{e.g.}, agentic reinforced evolution~\cite{fang2025comprehensive,dong2025agentic,gao2025survey}) to further improve collaborative capabilities.


\begin{table*}[t]
\centering
\caption{Comparison of Different Interactive Recommendation Agent Frameworks. The comparison criteria are categorized into three aspects: \colorbox{violet!12}{\textbf{Agent Capabilities}}, \colorbox{cyan!12}{\textbf{Interaction Features}}, and \colorbox{yellow!12}{\textbf{Practicality}}.}
\vspace{-1em}
\begin{tabular}{lccccc}
\toprule
\textbf{Capability} & \textbf{InteRecAgent}~\cite{huang2025recommender} & \textbf{RecBench+}~\cite{huang2025towards} & \textbf{InstructAgent}~\cite{xu2025instructagent} & \textbf{GOMMIR}~\cite{wu2023goal} & \textbf{RecBot (Ours)} \\
\midrule
\rowcolor{violet!12}
Collaborative Knowledge & \yes & \no & \no & \yes & \yes \\
\rowcolor{violet!12}
Tool Invocation & \yes & \no & \yes & \no & \yes \\
\rowcolor{violet!12}
Agent-Tuning & \yes & \no & \no & \yes & \yes \\
\rowcolor{violet!12}
Memory & \yes & \no & \yes & \yes & \yes \\
\rowcolor{violet!12}
Multi-Modal & \no & \no & \no & \yes & \yes \\
\addlinespace[0.2em]
\hline
\rowcolor{cyan!12}
Multi-turn Interaction & \yes & \no & \no & \yes & \yes \\
\rowcolor{cyan!12}
Complex User Command & \no & \no & \no & \no & \yes \\
\addlinespace[0.2em]
\hline
\rowcolor{yellow!12}
Industrial Deployment & \no & \no & \no & \no & \yes \\
\bottomrule
\end{tabular}
\label{tab:comparison}
\vspace{-1em}
\end{table*}

\subsection{Discussion}
In this section, we discuss RecBot's distinctions from existing interactive recommendation agents and the differences between Interactive Recommendation Feed (IRF) and Conversational Recommender systems (CRS) paradigms.
\subsubsection{Comparison of RecBot and Existing Interactive Recommendation Agents}
As summarized in Table~\ref{tab:comparison}, we compare RecBot against existing interactive recommendation agents across three dimensions: agent capabilities, interaction features, and practicality. RecBot demonstrates multi-aspect advancement as follows:
\begin{itemize}[leftmargin=*]
  \item \textbf{Agent Capabilities:} RecBot demonstrates five key advances: 
(1) \textit{Collaborative Knowledge}: Integrates historical user behaviors with explicit commands through intent-aware collaborative filtering.
(2) \textit{Tool Invocation}: Enables flexible scenario adaptation and seamless functionality extension through modular interfaces. 
(3) \textit{Agent-Tuning}: Achieves production-ready performance with cost-effective large language models via knowledge distillation. 
(4) \textit{Memory}: Maintains preference coherence across extended interactions through dynamic consolidation. 
(5) \textit{Multi-Modal}: Processes heterogeneous item representations across multiple modalities through unified encoding frameworks.
  \item \textbf{Interaction Features:} RecBot can handle complex multi-turn dialogues and sophisticated user commands (\textit{e.g.}, mixed positive and negative feedback, and interest shifts), addressing limitations of prior approaches that are constrained to single-turn or simplistic interactions, thereby enhancing real-world applicability.
  \item \textbf{Practical Impact:} RecBot achieves successful industrial deployment with validated business impact, while existing approaches remain largely theoretical or confined to academic prototypes.
\end{itemize}

\subsubsection{Interactive Recommendation Feed vs. Conversational Recommendation}
The IRF paradigm differs fundamentally from conversational recommender systems across two dimensions: 
\textbf{(1) Usage Scenario:} CRS address goal-directed search tasks where users possess well-defined information needs and explicit purchase intentions. IRF, conversely, facilitates exploratory browsing scenarios where users engage in open-ended discovery, with preferences emerging and evolving through iterative system interactions.
\textbf{(2) Preference Elicitation:} CRS employ structured interrogative approaches that strategically extract user preferences via predefined conversational schemas. IRF adopts a reactive paradigm where users provide contextually-anchored feedback through natural language responses to specific recommendation outputs, enabling preference articulation directly grounded in presented recommendation feed.

\section{Experiments}
In this section, we conduct extensive offline and online experiments to empirically demonstrate the effectiveness of the proposed RecBot framework. We first introduce the experimental setup, then present overall results and ablation studies, and finally provide in-depth analyses of online performance.

\subsection{Experimental Setup}
\subsubsection{\textbf{Datasets.}} 
We conduct our experiments on the following three recommendation datasets:
\begin{itemize}[leftmargin=*]
  \item \textbf{Amazon}~\cite{hou2024bridging}: We select the ``Books'' category from the Amazon dataset. Following previous studies~\cite{hou2025generating,tang2025think,zhou2020s3}, we treat all user historical reviews as interactions and apply 20-core filtering to remove users and items with fewer than 20 interactions, from which we sample 1,000 users for our experiments. For this dataset, we utilize price, language, and binding format as hard constraints, while book categories serve as soft conditions.
  \item \textbf{MovieLens}~\footnote{https://grouplens.org/datasets/movielens/}: We consider user-item interactions with ratings greater than 3 as positive interactions. For MovieLens, we employ movie release date as hard constraints, while movie genres as soft conditions for recommendation guidance.
  \item \textbf{Taobao}: The experiments utilize interaction records from Taobao, a leading e-commerce platform serving billions of users and items. The experimental scenario focuses on the ``\textit{Guess What You Like}'' column displayed on the Taobao APP's homepage, where the core task is to predict the next items users will interact with based on their historical behaviors and profile information. We sample 3,000 users for this offline experiment. In the Taobao dataset, we set price, style, and material as hard constraints, while the platform's multi-level product categories serve as soft conditions. Additionally, we incorporate product cover images as visual features to enhance multimodal understanding.
\end{itemize}

Consistent with existing works~\cite{huang2025recommender}, we adopt the commonly-used Leave-One-Out (LOO) evaluation protocol, where user historical interactions are organized in chronological order, and the last item in each user sequence is held out for testing.

\subsubsection{\textbf{Offline Interactive Recommendation Experiments.}}
To closely simulate realistic online user-system interaction environments, we employ GPT-4.1\footnote{The specific snapshot version used is GPT-4.1-2025-04-14.} as the backbone for our user simulation framework, leveraging its robust language generation capabilities to emulate authentic user behaviors. Based on prevalent user command patterns observed in production systems, we design three distinct interaction scenarios of increasing complexity:
\begin{itemize}[leftmargin=*]
    \item \textbf{Single-Round Interaction (SR):} Users possess clearly defined preferences and can express their requirements for target item characteristics comprehensively within a single command. This scenario is suited for users who have clear intentions.
    \item \textbf{Multi-Round Interaction (MR):} Users initially exhibit ambiguous or exploratory preferences. Through iterative exposure to recommendation feeds, users progressively refine their requirements by providing both positive interest signals for desired attributes and negative feedback regarding unsatisfactory aspects of current recommendations. In our implementation, we employ pre-constructed prompting strategies to drive diverse user feedback styles, enabling both proactive new requirements and reactive negative feedback to exposed items. This iterative refinement process expects the system to dynamically adjust its recommendation policy based on accumulated user guidance.
    \item \textbf{Multi-Round Interaction with Interest Drift (MRID):} Building upon the MR scenario, this setting introduces preference evolution where users may exhibit conflicting or shifting interests across interaction rounds. For instance, a user initially seeking Windows-based computers may subsequently pivot toward Mac preferences during the browsing session. This scenario represents the most challenging yet realistic user behavior pattern commonly observed in exploratory contexts. In our implementation, we simulate this interest drift by initially pre-sampling a random item as a pseudo-target to guide the user's early interactions, then strategically redirecting the user's preference toward their actual ground-truth next interaction item at round 3, thereby creating a realistic preference shift scenario.
\end{itemize}

For multi-round interaction scenarios (\textit{i.e.}, MR and MRID), we establish experimental parameters with a maximum interaction limit of 5 rounds and recommendation list size of $K=5$ items per round, reflecting typical user attention spans focused on top-ranked recommendations~\cite{jeunen2023probabilistic}. We assume that users will select the top-1 recommended item from each exposed list to issue explicit feedback commands that guide the system's strategy adjustment for subsequent rounds. The interaction terminates when either: (1) the user's ground-truth target item appears within the top-5 recommendations, indicating successful preference satisfaction, or (2) the maximum round limit is exceeded, representing interaction failure or user abandonment.
These different experimental designs enable systematic evaluation of RecBot's adaptive capabilities across diverse user interaction patterns while maintaining computational feasibility and realistic user behavior modeling.

\subsubsection{\textbf{Evaluation Metrics.}}
For offline experiments, we follow prior work~\cite{wang2022user,wang2024can} and adopt standard ranking-based metrics including \textbf{Recall@N} and \textbf{NDCG@N}, where $N \in \{10, 20, 50\}$. Additionally, we introduce three specialized metrics to comprehensively evaluate interactive recommendation performance:
\begin{itemize}[leftmargin=*]
  \item \textbf{Condition Satisfaction Rate (CSR@N):} This metric measures the attribute-level coverage ratio between recommended items in top-$N$ positions and the target item's characteristics. CSR@N serves as an attribute-oriented ranking accuracy measure that, compared to the strict precision requirements of Recall and NDCG, reflects the model's capability to correctly infer user preferences regarding item category and attribute preferences.
  \item \textbf{Pass Rate (PR):} This binary metric calculates whether the target item is successfully delivered within the top-$K$ recommendation feed during the limited interaction rounds $T$ ($T{=}1$ for SR scenarios, $T{=}5$ for MR and MRID scenarios).
  \item \textbf{Average Rounds (AR):} This efficiency metric quantifies the average interaction rounds required for the system to successfully predict the target item to users at the top-$K$ positions. For cases where the system fails to recommend the target item within the maximum $T$ rounds, we assign a penalty score of $T{+}1$ to reflect unsuccessful interactions.
\end{itemize}

\emph{Note:} Among all offline metrics, AR follows a ``lower-is-better'' principle, indicating that the system achieves higher efficiency by identifying correct target items through fewer interaction rounds. All other metrics follow a ``higher-is-better'' evaluation paradigm.

For online experiments, we assess RecBot's real-world performance through two complementary evaluation dimensions that capture both user satisfaction and business impact.
In specific, \underline{\textbf{(1) User Experience Metrics}} focus on measuring user satisfaction and engagement quality, including: \textbf{Negative Feedback Frequency (NFF)}, measuring the frequency of user dissatisfaction signals toward recommendation feeds; \textbf{Exposed Item Category Diversity (EICD)}, quantifying the average category diversity of items presented to users; \textbf{Clicked Item Category Diversity (CICD)}, measuring the average category diversity of items users actively engage with through clicks.
\underline{\textbf{(2) Business Utility Metrics}} evaluate the commercial utility, including \textbf{Page Views (PV)}, capturing overall user engagement with recommended content; \textbf{Add-to-Cart (ATC)}, measuring conversion intent through shopping cart additions; \textbf{Gross Merchandise Volume (GMV)}, quantifying the total transaction value generated from recommendations.

\emph{Note.} Among the above online metrics, NFF follows a ``lower-is-better'' evaluation criterion, as reduced negative feedback frequency indicates improved user satisfaction with recommendation quality. All remaining online metrics adhere to a ``higher-is-better'' assessment protocol, where increased values demonstrate enhanced user engagement and business benefits.

\begin{table*}[t]
\centering
\caption{Performance comparison of different methods on Single-Round (SR) interaction scenarios across three datasets. We report results based on Recall@K (R@K), NDCG@K (N@K), Condition Satisfaction Rate@K (C@K), and Pass Rate (PR) metrics. Best results are highlighted in bold.}
\label{tab:main_offline_st_table}
\begin{tabular}{l|ccc|ccc|ccc|cc}
\toprule
\multicolumn{11}{c}{\textbf{Amazon (SR)}} \\
\midrule
\textbf{Method} & \textbf{R@10} & \textbf{R@20} & \textbf{R@50} & \textbf{N@10} & \textbf{N@20} & \textbf{N@50} & \textbf{C@10} & \textbf{C@20} & \textbf{C@50} & \textbf{PR} \\
\midrule
\cellcolor{white}SASRec & 0.0098 & 0.0163 & 0.0326 & 0.0061 & 0.0077 & 0.0109 & 65.81\% & 69.08\% & 73.88\% & 0.76\% \\
\cellcolor{white}BERT4Rec & 0.0076 & 0.0109 & 0.0229 & 0.0054 & 0.0062 & 0.0085 & 64.84\% & 69.40\% & 73.82\% & 0.65\% \\
\cellcolor{white}MoRec & 0.0316 & 0.0457 & 0.0827 & 0.0184 & 0.0220 & 0.0292 & 73.49\% & 77.65\% & 82.46\% & 2.29\% \\
\cellcolor{white}UniSRec & 0.0370 & 0.0544 & 0.1001 & 0.0202 & 0.0246 & 0.0336 & 73.87\% & 77.65\% & 81.99\% & 2.18\% \\
\hline
\cellcolor{white}BM25 & 0.0283 & 0.0370 & 0.0816 & 0.0128 & 0.0150 & 0.0239 & 80.51\% & 85.96\% & 91.29\% & 1.31\% \\
\cellcolor{white}BGE & 0.0598 & 0.1012 & 0.1795 & 0.0284 & 0.0387 & 0.0543 & 92.76\% & 95.19\% & 97.23\% & 3.16\% \\
\hline
\cellcolor{white}GOMMIR & 0.0011 & 0.0022 & 0.0054 & 0.0004 & 0.0007 & 0.0013 & 53.14\% & 60.89\% & 71.65\% & 0.00\% \\
\cellcolor{white}InteRecAgent & 0.0609 & 0.1023 & 0.1708 & 0.0321 & 0.0425 & 0.0560 & 81.98\% & 84.21\% & 87.28\% & 3.81\% \\
\cellcolor{white}Intruct2Agent & 0.0033 & 0.0054 & 0.0131 & 0.0011 & 0.0016 & 0.0031 & 67.02\% & 72.42\% & 78.53\% & 0.00\% \\
\hline
\cellcolor{white}RecBot-Qwen (Orig.) & 0.2078 & 0.2851 & 0.4091 & 0.1119 & 0.1313 & 0.1555 & 92.12\% & 93.70\% & 95.02\% & 14.04\% \\
\cellcolor{white}RecBot-Qwen (Align.) & 0.2198 & 0.3101 & 0.4614 & 0.1207 & 0.1434 & 0.1736 & 94.92\% & 96.13\% & 97.34\% & 14.58\% \\
\cellcolor{white}RecBot-GPT & \textbf{0.2459} & \textbf{0.3547} & \textbf{0.5212} & \textbf{0.1391} & \textbf{0.1667} & \textbf{0.1994} & \textbf{97.63\%} & \textbf{98.47\%} & \textbf{99.39\%} & \textbf{16.76\%} \\
\midrule
\multicolumn{11}{c}{\textbf{MovieLens (SR)}} \\
\midrule
\cellcolor{white}SASRec & 0.0236 & 0.0428 & 0.0942 & 0.0130 & 0.0177 & 0.0279 & 57.55\% & 62.15\% & 70.45\% & 1.61\% \\
\cellcolor{white}BERT4Rec & 0.0364 & 0.0493 & 0.0996 & 0.0179 & 0.0212 & 0.0311 & 57.44\% & 61.99\% & 70.24\% & 1.71\% \\
\cellcolor{white}MoRec & 0.0642 & 0.1103 & 0.2366 & 0.0302 & 0.0418 & 0.0663 & 60.65\% & 66.43\% & 76.61\% & 3.00\% \\
\cellcolor{white}UniSRec & 0.0664 & 0.1338 & 0.2612 & 0.0368 & 0.0539 & 0.0791 & 60.28\% & 66.60\% & 76.71\% & 4.07\% \\
\hline
\cellcolor{white}BM25 & 0.0257 & 0.0535 & 0.1381 & 0.0123 & 0.0192 & 0.0358 & 60.81\% & 66.86\% & 77.19\% & 1.28\% \\
\cellcolor{white}BGE & 0.1370 & 0.2077 & 0.3512 & 0.0645 & 0.0822 & 0.1101 & 77.25\% & 85.06\% & 92.24\% & 7.49\% \\
\hline
\cellcolor{white}GOMMIR & 0.0021 & 0.0043 & 0.0150 & 0.0007 & 0.0013 & 0.0033 & 43.36\% & 43.36\% & 61.03\% & 0.11\% \\
\cellcolor{white}InteRecAgent & 0.1103 & 0.1959 & 0.3555 & 0.0591 & 0.0807 & 0.0591 & 66.81\% & 73.45\% & 82.92\% & 6.64\% \\
\cellcolor{white}Intruct2Agent & 0.0214 & 0.0460 & 0.0835 & 0.0087 & 0.0148 & 0.0224 & 54.12\% & 59.37\% & 68.58\% & 1.28\% \\
\hline
\cellcolor{white}RecBot-Qwen (Orig.) & 0.3073 & 0.4058 & 0.5589 & 0.1935 & 0.2181 & 0.2484 & 80.57\% & 83.99\% & 88.22\% & 23.13\% \\
\cellcolor{white}RecBot-Qwen (Align.) & 0.3383 & 0.4208 & 0.5675 & 0.2101 & 0.2309 & 0.2598 & 81.10\% & 84.69\% & 88.38\% & 25.70\% \\
\cellcolor{white}RecBot-GPT & \textbf{0.4293} & \textbf{0.5161} & \textbf{0.6649} & \textbf{0.2749} & \textbf{0.2967} & \textbf{0.3264} & \textbf{88.81\%} & \textbf{91.22\%} & \textbf{93.36\%} & \textbf{32.76\%} \\
\midrule
\multicolumn{11}{c}{\textbf{Taobao (SR)}} \\
\midrule
\cellcolor{white}SASRec & 0.0267 & 0.0400 & 0.0744 & 0.0140 & 0.0173 & 0.0240 & 37.75\% & 40.36\% & 44.71\% & 1.67\% \\
\cellcolor{white}BERT4Rec & 0.0210 & 0.0340 & 0.0654 & 0.0101 & 0.0134 & 0.0196 & 36.56\% & 39.74\% & 44.23\% & 1.20\% \\
\cellcolor{white}MoRec & 0.1421 & 0.1922 & 0.2639 & 0.0703 & 0.0830 & 0.0972 & 45.31\% & 49.13\% & 54.67\% & 9.48\% \\
\cellcolor{white}UniSRec & 0.1535 & 0.2095 & 0.2840 & 0.0734 & 0.0875 & 0.1023 & 45.16\% & 49.14\% & 54.21\% & 9.41\% \\
\hline
\cellcolor{white}BM25 & 0.0000 & 0.0000 & 0.0000 & 0.0000 & 0.0000 & 0.0000 & 28.10\% & 34.00\% & 41.38\% & 0.00\% \\
\cellcolor{white}BGE & 0.2025 & 0.2753 & 0.3947 & 0.1174 & 0.1358 & 0.1594 & 82.60\% & 86.35\% & 90.22\% & 14.25\% \\
\hline
\cellcolor{white}GOMMIR & 0.0003 & 0.0003 & 0.0007 & 0.0001 & 0.0001 & 0.0002 & 17.24\% & 19.70\% & 23.49\% & 0.03\% \\
\cellcolor{white}InteRecAgent & 0.2122 & 0.2673 & 0.3347 & 0.1184 & 0.1323 & 0.1457 & 57.18\% & 60.74\% & 65.89\% & 15.72\% \\
\cellcolor{white}Intruct2Agent & 0.0230 & 0.0377 & 0.0704 & 0.0104 & 0.0140 & 0.0205 & 34.15\% & 41.11\% & 50.10\% & 1.20\% \\
\hline
\cellcolor{white}RecBot-Qwen (Orig.) & 0.3797 & 0.4668 & 0.5756 & 0.2432 & 0.2650 & 0.2866 & 81.18\% & 84.59\% & 88.13\% & 29.56\% \\
\cellcolor{white}RecBot-Qwen (Align.) & 0.4531 & 0.5402 & 0.6483 & 0.2875 & 0.3094 & 0.3310 & 86.12\% & 88.39\% & 91.02\% & 35.37\% \\
\cellcolor{white}RecBot-GPT & \textbf{0.4918} & \textbf{0.5879} & \textbf{0.6980} & \textbf{0.3174} & \textbf{0.3417} & \textbf{0.3636} & \textbf{89.40\%} & \textbf{91.58\%} & \textbf{93.83\%} & \textbf{39.31\%} \\
\bottomrule
\end{tabular}
\end{table*}

\begin{table*}[t]
\centering
\caption{Performance comparison on Multi-Round (MR) interaction scenarios. We report results based on Recall (R@K), NDCG (N@K), Condition Satisfaction Rate (C@K), Pass Rate (PR), and Average Rounds (AR) metrics. Best results are highlighted in bold. ``$\uparrow$'' indicates that higher values are better, while ``$\downarrow$'' indicates that lower values are better.}
\label{tab:main_offline_mt_table}
\resizebox{\textwidth}{!}{%
\begin{tabular}{l|ccc|ccc|ccc|cc}
\toprule
\multicolumn{12}{c}{\textbf{Amazon (MR)}} \\
\midrule
\textbf{Method} & \textbf{R@10}$\bm{\uparrow}$ & \textbf{R@20}$\bm{\uparrow}$ & \textbf{R@50}$\bm{\uparrow}$ & \textbf{N@10}$\bm{\uparrow}$ & \textbf{N@20}$\bm{\uparrow}$ & \textbf{N@50}$\bm{\uparrow}$ & \textbf{C@10}$\bm{\uparrow}$ & \textbf{C@20}$\bm{\uparrow}$ & \textbf{C@50}$\bm{\uparrow}$ & \textbf{PR}$\bm{\uparrow}$ & \textbf{AR}$\bm{\downarrow}$ \\
\midrule
\cellcolor{white}BM25 & 0.0381 & 0.0479 & 0.0664 & 0.0253 & 0.0277 & 0.0313 & 77.37\% & 83.44\% & 88.26\% & 3.70\% & 5.8194 \\
\cellcolor{white}BGE & 0.0609 & 0.0740 & 0.1153 & 0.0329 & 0.0361 & 0.0444 & 87.29\% & 89.84\% & 93.05\% & 5.22\% & 5.7661 \\
\hline
\cellcolor{white}GOMMIR & 0.0000 & 0.0011 & 0.0033 & 0.0000 & 0.0003 & 0.0007 & 52.20\% & 64.01\% & 71.21\% & 0.00\% & 6.0000 \\
\cellcolor{white}InteRecAgent & 0.0533 & 0.0849 & 0.1436 & 0.0284 & 0.0363 & 0.0479 & 78.89\% & 81.77\% & 85.24\% & 3.70\% & 5.8085 \\
\cellcolor{white}Intruct2Agent & 0.0250 & 0.0283 & 0.0326 & 0.0170 & 0.0177 & 0.0187 & 66.69\% & 71.98\% & 77.35\% & 2.50\% & 5.8596 \\
\hline
\cellcolor{white}RecBot-Qwen (Orig.) & 0.1632 & 0.1904 & 0.2677 & 0.0882 & 0.0951 & 0.1104 & 82.30\% & 83.99\% & 86.43\% & 12.30\% & 5.6126 \\
\cellcolor{white}RecBot-Qwen (Align.) & 0.1893 & 0.2416 & 0.3177 & 0.1044 & 0.1177 & 0.1327 & 84.21\% & 86.20\% & 88.71\% & 15.02\% & 5.5234 \\
\cellcolor{white}RecBot-GPT & \textbf{0.2296} & \textbf{0.3199} & \textbf{0.4374} & \textbf{0.1250} & \textbf{0.1475} & \textbf{0.1708} & \textbf{95.20\%} & \textbf{96.56\%} & \textbf{97.90\%} & \textbf{17.19\%} & \textbf{5.4244} \\
\midrule
\multicolumn{12}{c}{\textbf{MovieLens (MR)}} \\
\midrule
\cellcolor{white}BM25 & 0.0621 & 0.0931 & 0.2034 & 0.0381 & 0.0458 & 0.0673 & 63.28\% & 68.47\% & 78.80\% & 5.57\% & 5.7099 \\
\cellcolor{white}BGE & 0.1146 & 0.1745 & 0.3266 & 0.0617 & 0.0765 & 0.1063 & 71.63\% & 79.76\% & 89.61\% & 9.53\% & 5.5675 \\
\hline
\cellcolor{white}GOMMIR & 0.0043 & 0.0064 & 0.0171 & 0.0018 & 0.0024 & 0.0044 & 43.47\% & 43.47\% & 61.03\% & 0.32\% & 5.9818 \\
\cellcolor{white}InteRecAgent & 0.1081 & 0.1820 & 0.3255 & 0.0582 & 0.0767 & 0.1047 & 65.42\% & 71.31\% & 80.89\% & 7.49\% & 5.6006 \\
\cellcolor{white}Intruct2Agent & 0.0557 & 0.0749 & 0.1221 & 0.0351 & 0.0398 & 0.0489 & 55.09\% & 60.55\% & 67.56\% & 5.03\% & 5.7173 \\
\hline
\cellcolor{white}RecBot-Qwen (Orig.) & 0.3812 & 0.4625 & 0.5717 & 0.2407 & 0.2610 & 0.2828 & 84.21\% & 86.30\% & 89.08\% & 34.05\% & 4.9261 \\
\cellcolor{white}RecBot-Qwen (Align.) & \textbf{0.4315} & \textbf{0.5021} & \textbf{0.6221} & \textbf{0.2742} & \textbf{0.2921} & \textbf{0.3159} & \textbf{87.53\%} & \textbf{89.61\%} & \textbf{92.51\%} & \textbf{38.12\%} & \textbf{4.7837} \\
\cellcolor{white}RecBot-GPT & 0.4036 & 0.4829 & 0.6146 & 0.2549 & 0.2749 & 0.3013 & 86.13\% & 88.28\% & 90.79\% & 35.01\% & 4.8298 \\
\midrule
\multicolumn{12}{c}{\textbf{Taobao (MR)}} \\
\midrule
\cellcolor{white}BM25 & 0.0941 & 0.0941 & 0.0941 & 0.0540 & 0.0540 & 0.0540 & 35.01\% & 40.33\% & 47.01\% & 9.41\% & 5.4354 \\
\cellcolor{white}BGE & 0.1919 & 0.2259 & 0.2880 & 0.1059 & 0.1143 & 0.1267 & 64.11\% & 68.30\% & 74.10\% & 17.18\% & 5.1235 \\
\hline
\cellcolor{white}GOMMIR & 0.0000 & 0.0000 & 0.0003 & 0.0000 & 0.0000 & 0.0001 & 15.04\% & 19.07\% & 23.11\% & 0.00\% & 6.0000 \\
\cellcolor{white}InteRecAgent & 0.2166 & 0.2619 & 0.3160 & 0.1137 & 0.1252 & 0.1360 & 53.26\% & 57.21\% & 62.49\% & 18.42\% & 5.0791 \\
\cellcolor{white}Intruct2Agent & 0.1061 & 0.1091 & 0.1205 & 0.0602 & 0.0610 & 0.0632 & 33.85\% & 38.67\% & 46.05\% & 10.44\% & 5.3971 \\
\hline
\cellcolor{white}RecBot-Qwen (Orig.) & 0.3967 & 0.4518 & 0.5199 & 0.2415 & 0.2553 & 0.2689 & 82.44\% & 84.83\% & 87.76\% & 35.14\% & 4.5122 \\
\cellcolor{white}RecBot-Qwen (Align.) & 0.4238 & 0.4735 & 0.5252 & 0.2627 & 0.2752 & 0.2855 & 76.94\% & 79.35\% & 81.98\% & 38.47\% & 4.3827 \\
\cellcolor{white}RecBot-GPT & \textbf{0.4618} & \textbf{0.5305} & \textbf{0.6220} & \textbf{0.2795} & \textbf{0.2967} & \textbf{0.3149} & \textbf{84.86\%} & \textbf{87.32\%} & \textbf{90.22\%} & \textbf{41.14\%} & \textbf{4.2809} \\
\bottomrule
\end{tabular}
}
\end{table*}

\begin{table*}[t]
\centering
\caption{Performance comparison on Multi-Round with Interest Drift (MRID) scenarios.  We report results based on Recall (R@K), NDCG (N@K), Condition Satisfaction Rate (C@K), Pass Rate (PR), and Average Rounds (AR) metrics. Best results are highlighted in bold. ``$\uparrow$'' indicates that higher values are better, while ``$\downarrow$'' indicates that lower values are better.}
\label{tab:main_offline_mtid_table}
\resizebox{\textwidth}{!}{%
\begin{tabular}{l|ccc|ccc|ccc|cc}
\toprule
\multicolumn{12}{c}{\textbf{Amazon (MRID)}} \\
\midrule
\textbf{Method} & \textbf{R@10}$\bm{\uparrow}$ & \textbf{R@20}$\bm{\uparrow}$ & \textbf{R@50}$\bm{\uparrow}$ & \textbf{N@10}$\bm{\uparrow}$ & \textbf{N@20}$\bm{\uparrow}$ & \textbf{N@50}$\bm{\uparrow}$ & \textbf{C@10}$\bm{\uparrow}$ & \textbf{C@20}$\bm{\uparrow}$ & \textbf{C@50}$\bm{\uparrow}$ & \textbf{PR}$\bm{\uparrow}$ & \textbf{AR}$\bm{\downarrow}$ \\
\midrule
\cellcolor{white}BM25 & 0.0424 & 0.0544 & 0.0860 & 0.0241 & 0.0270 & 0.0332 & 78.55\% & 84.28\% & 89.55\% & 3.16\% & 5.8564 \\
\cellcolor{white}BGE & 0.0316 & 0.0392 & 0.0555 & 0.0198 & 0.0216 & 0.0248 & 80.65\% & 85.14\% & 89.59\% & 2.72\% & 5.8531 \\
\hline
\cellcolor{white}GOMMIR & 0.0011 & 0.0011 & 0.0033 & 0.0004 & 0.0004 & 0.0008 & 50.08\% & 60.05\% & 65.44\% & 0.00\% & 6.0000 \\
\cellcolor{white}InteRecAgent & 0.0511 & 0.0783 & 0.1273 & 0.0263 & 0.0333 & 0.0429 & 76.48\% & 79.67\% & 83.44\% & 3.26\% & 5.8335 \\
\cellcolor{white}Intruct2Agent & 0.0239 & 0.0250 & 0.0272 & 0.0165 & 0.0167 & 0.0172 & 65.26\% & 70.97\% & 76.69\% & 2.29\% & 5.8640 \\
\hline
\cellcolor{white}RecBot-Qwen (Orig.) & 0.0958 & 0.1360 & 0.2133 & 0.0507 & 0.0607 & 0.0758 & 77.54\% & 81.01\% & 84.70\% & 6.86\% & 5.8466 \\
\cellcolor{white}RecBot-Qwen (Align.) & 0.1056 & 0.1371 & 0.1937 & 0.0595 & 0.0673 & 0.0784 & 77.84\% & 80.73\% & 85.07\% & 8.38\% & 5.8292 \\
\cellcolor{white}RecBot-GPT & \textbf{0.1164} & \textbf{0.1513} & \textbf{0.2688} & \textbf{0.0651} & \textbf{0.0740} & \textbf{0.0972} & \textbf{89.13\%} & \textbf{91.68\%} & \textbf{93.51\%} & \textbf{8.49\%} & \textbf{5.8324} \\
\midrule
\multicolumn{12}{c}{\textbf{MovieLens (MRID)}} \\
\midrule
\cellcolor{white}BM25 & 0.0600 & 0.0942 & 0.2120 & 0.0365 & 0.0449 & 0.0679 & 62.31\% & 69.49\% & 80.57\% & 5.03\% & 5.7195 \\
\cellcolor{white}BGE & 0.0974 & 0.1360 & 0.2463 & 0.0551 & 0.0646 & 0.0864 & 66.33\% & 73.77\% & 82.66\% & 6.64\% & 5.6895 \\
\hline
\cellcolor{white}GOMMIR & 0.0043 & 0.0064 & 0.0171 & 0.0018 & 0.0024 & 0.0044 & 43.47\% & 43.47\% & 61.03\% & 0.32\% & 5.9818 \\
\cellcolor{white}InteRecAgent & 0.1039 & 0.1788 & 0.3191 & 0.0554 & 0.0743 & 0.1018 & 65.90\% & 71.73\% & 80.94\% & 7.17\% & 5.6702 \\
\cellcolor{white}Intruct2Agent & 0.0578 & 0.0696 & 0.1028 & 0.0375 & 0.0404 & 0.0468 & 54.93\% & 58.94\% & 67.56\% & 5.35\% & 5.7099 \\
\hline
\cellcolor{white}RecBot-Qwen (Orig.) & 0.2955 & 0.3683 & 0.4743 & 0.1870 & 0.2055 & 0.2268 & 81.96\% & 84.53\% & 88.06\% & 24.52\% & 5.4111 \\
\cellcolor{white}RecBot-Qwen (Align.) & \textbf{0.3940} & \textbf{0.4582} & \textbf{0.5867} & \textbf{0.2506} & \textbf{0.2668} & \textbf{0.2922} & \textbf{84.05\%} & \textbf{85.92\%} & \textbf{88.60\%} & \textbf{33.51\%} & \textbf{5.2591} \\
\cellcolor{white}RecBot-GPT & 0.3158 & 0.4111 & 0.5675 & 0.1945 & 0.2183 & 0.2491 & 81.48\% & 84.85\% & 89.19\% & 26.02\% & 5.3951 \\
\midrule
\multicolumn{12}{c}{\textbf{Taobao (MRID)}} \\
\midrule
\cellcolor{white}BM25 & 0.0941 & 0.0941 & 0.0941 & 0.0540 & 0.0540 & 0.0540 & 36.25\% & 41.32\% & 47.47\% & 9.41\% & 5.4354 \\
\cellcolor{white}BGE & 0.0998 & 0.1024 & 0.1091 & 0.0561 & 0.0568 & 0.0581 & 37.41\% & 42.32\% & 49.16\% & 9.68\% & 5.4308 \\
\hline
\cellcolor{white}GOMMIR & 0.0000 & 0.0000 & 0.0003 & 0.0000 & 0.0000 & 0.0001 & 14.93\% & 19.25\% & 23.07\% & 0.00\% & 6.0000 \\
\cellcolor{white}InteRecAgent & 0.2089 & 0.2526 & 0.3103 & 0.1085 & 0.1196 & 0.1312 & 51.94\% & 55.52\% & 60.41\% & 16.85\% & 5.2155 \\
\cellcolor{white}Intruct2Agent & 0.0948 & 0.0948 & 0.0948 & 0.0543 & 0.0543 & 0.0543 & 23.55\% & 26.45\% & 29.99\% & 9.44\% & 5.4348 \\
\hline
\cellcolor{white}RecBot-Qwen (Orig.) & 0.2396 & 0.2736 & 0.3203 & 0.1335 & 0.1422 & 0.1514 & 56.50\% & 59.14\% & 62.68\% & 20.42\% & 5.2456 \\
\cellcolor{white}RecBot-Qwen (Align.) & \textbf{0.2964} & \textbf{0.3432} & 0.3953 & \textbf{0.1793} & \textbf{0.1911} & \textbf{0.2016} & 65.00\% & 67.80\% & 71.10\% & \textbf{25.60\%} & 5.1906 \\
\cellcolor{white}RecBot-GPT & 0.2826 & 0.3377 & \textbf{0.4141} & 0.1622 & 0.1761 & 0.1914 & \textbf{72.18\%} & \textbf{75.34\%} & \textbf{79.28\%} & 22.92\% & \textbf{5.1869} \\
\bottomrule
\end{tabular}
}
\end{table*}

\subsubsection{\textbf{Baselines.}}
We compare RecBot against various traditional sequential recommendation algorithms, instruction-aware methods, and interactive recommendation agent approaches as follows:

\underline{\textbf{(1) Traditional Sequential Recommendation Methods:}}
\begin{itemize}[leftmargin=*]
  \item \textbf{SASRec}~\cite{kang2018self}, a self-attention based sequential recommendation model that adaptively weights user historical interactions to predict the next item.
  \item \textbf{BERT4Rec}~\cite{sun2019bert4rec}, applies bidirectional modeling to user behavior sequences, using masked item prediction with self-attention to leverage both past and future context information.
  \item \textbf{MoRec}~\cite{yuan2023go}, a modality-based recommendation approach that replaces ID embeddings with pre-trained encoders through end-to-end training to leverage item content features.
  \item \textbf{UniSRec}~\cite{hou2022towards}, a universal sequence representation learning approach that leverages item text descriptions with parametric whitening and mixture-of-experts enhanced adaptors to learn transferable representations across domains.
\end{itemize}

\underbar{\textbf{(2) Command-Aware Recommendation Methods:}}
\begin{itemize}[leftmargin=*]
  \item \textbf{BM25}~\cite{robertson2009probabilistic}, a classic probabilistic ranking function for information retrieval that computes relevance scores between user queries and item descriptions using term frequency and inverse document frequency with length normalization parameters.
  \item \textbf{BGE}~\cite{xiao2024c}, a ranking-based recommendation approach that computes item relevance scores using BGE embeddings to measure semantic similarity between user queries and candidate items for recommendation ranking.
\end{itemize}

\underbar{\textbf{(3) Interactive Recommendation Agent Methods:}}
\begin{itemize}[leftmargin=*]
  \item \textbf{GOMMIR}~\cite{wu2023goal}, a goal-oriented multi-modal interactive recommendation model that addresses the coupling issue between policy optimization and representation learning through joint supervised and reinforcement learning.
  \item \textbf{InteRecAgent}~\cite{huang2025recommender}, a LLM-based interactive recommender framework that bridges traditional recommendation models with large language models using plan-first execution, dynamic demonstrations, and reflection mechanisms with shared candidate bus.
  \item \textbf{Instruct2Agent}~\cite{xu2025instructagent}, a user-controllable recommender agent that serves as a protective interface between users and recommender systems using instruction-aware parsing, external knowledge tools, and dynamic memory mechanisms.
  \item \textbf{RecBot (Ours)}, to evaluate our proposed RecBot, we conduct experiments across three distinct configurations as follows:
  \textbf{(1) RecBot-Qwen (Orig.)}, serves as our base version utilizing the Qwen3-14B~\cite{yang2025qwen3} foundation model.
  \textbf{(2) RecBot-Qwen (Align.)} is an enhanced variant built upon the Qwen3-14B foundation model that integrates the multi-agent optimization strategy detailed in Sec.~\ref{sec:optimization}, where GPT-4.1 serves as the Teacher RecBot Agent to guide the simulation-augmented knowledge distillation.
  \textbf{(3) RecBot-GPT}, leverages the advanced GPT-4.1~\cite{achiam2023gpt} to demonstrate the performance potential of our framework when powered by state-of-the-art closed-source models.
\end{itemize}

\subsubsection{\textbf{Implementation Details.}}
For multi-agent optimization, we employ Qwen3-14B-Instruct~\cite{yang2025qwen3} as our backbone large language model, implemented using OpenRLHF~\cite{hu2024openrlhf} and PyTorch~\cite{paszke2019pytorch} frameworks on 8$\times$NVIDIA A100 GPUs. We adopt LoRA~\cite{hu2022lora} for Parameter-Efficient Fine-Tuning (PEFT) with rank=32, alpha=32, dropout=0.1, training for 3 epochs with learning rate $5e^{-6}$ and batch size 256. User historical interaction sequences are truncated to a maximum length of 50 items, and BGE~\cite{xiao2024c} serves as the pre-trained embedding model for all semantic computations. The simulation-augmented knowledge distillation leverages GPT-4.1~\cite{achiam2023gpt} as the teacher model to generate training trajectories for fine-tuning.

\subsection{Offline Experiments}

\subsubsection{Overall Performance}
The performance comparison results across Single-Round (SR), Multi-Round (MR), and Multi-Round with Interest Drift (MRID) interaction scenarios are presented in Tables~\ref{tab:main_offline_st_table},~\ref{tab:main_offline_mt_table}, and~\ref{tab:main_offline_mtid_table}, respectively. These results reveal several compelling findings as follows:

(1) Compared to ID-based user interaction modeling, multimodal fusion approaches incorporating textual (and visual) information more effectively capture users' fine-grained sequence evolving patterns. Across all datasets and metrics, MoRec and UniSRec significantly outperform SASRec and BERT4Rec. On the Amazon SR task, for example, UniSRec delivers a Recall@10 of 0.0370, markedly surpassing SASRec (0.0098). This demonstrates the critical importance of leveraging rich content representations beyond simple ID embeddings for nuanced preference modeling.

(2) While traditional recommendation models rely on implicit behavioral signals for preference inference, their inability to incorporate users' explicit feedback fundamentally limits timely policy adjustment based on linguistic commands. Command-aware methods achieve superior performance by explicitly matching user instructions with semantically relevant items. Moreover, BGE consistently outperforms BM25 in most cases due to its pre-training on extensive text matching corpora, which facilitates deep semantic comprehension that transcends BM25's shallow token-level matching. This performance gap is particularly evident on MovieLens SR, where BGE obtains 0.1370 Recall@10 versus BM25's 0.0257.

(3) Existing interactive recommendation agents reveal fundamental architectural limitations in complex scenarios. In specific, GOMMIR exhibits catastrophic failure with near-zero performance due to its overly naive encoding approach through shallow transformers. While InteRec2Agent incorporates most advanced agent design modules including tool invocation and memory mechanisms, it still lacks fine-grained decomposition and parsing capabilities for user commands, hindering its performance in complex interaction scenarios and lacking command-aware multimodal sequential modeling abilities. InstructAgent demonstrates limited capability due to heavy reliance on LLMs' intrinsic abilities while lacking domain-specific collaborative knowledge.

(4) Our RecBot achieves state-of-the-art performance across all datasets, tasks and metrics, with particularly remarkable efficiency in multi-round scenarios. For example, on Taobao MR, RecBot-GPT achieves 41.14\% pass rate with 4.2809 average rounds compared to the InteRecAgent at 18.42\% pass rate with 5.0791 rounds. Most intriguingly, our knowledge distillation-based mulit-agent optimization enables student models to surpass their teachers. Specifically, RecBot-Qwen (Align.) achieves 0.3940 Recall@10 and 33.51\% pass rate on MovieLens MRID, exceeding its GPT-4.1 teacher's 0.3158 Recall@10 and 26.02\% pass rate. This phenomenon aligns with previous work~\cite{hsieh2023distilling,xu2024slmrec} demonstrating that focused knowledge transfer can unlock latent capabilities in smaller models, providing both practical deployment advantages and theoretical insights into the untapped potential of student model.

\subsubsection{Ablation Study}

\begin{figure}[t]
\centering
\includegraphics[width=\linewidth]{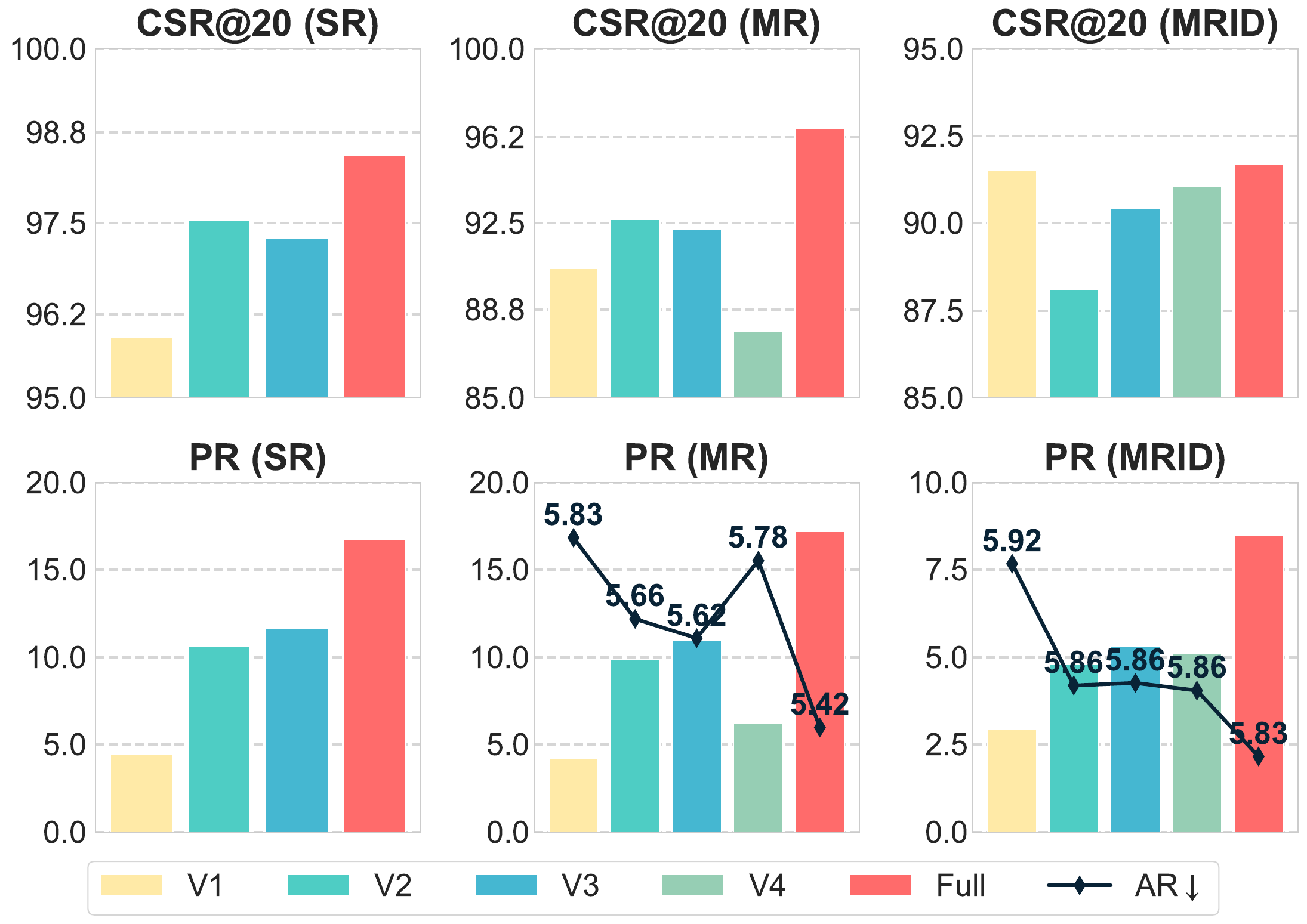}
\caption{Offline ablation study results on Amazon dataset. All numerical values on axes correspond to percentages (percentage notation is omitted for conciseness).}
\vspace{-2em}
\label{fig:offline_ablation}
\end{figure}

To further evaluate the contribution of each component within RecBot, we conduct comprehensive ablation experiments by comparing several simplified variants against the complete framework. Specifically, we examine five configurations: \textbf{(1) V1} employs only the Semantic Relevance Path of the Matcher Tool, relying exclusively on semantic similarity between structured positive user preferences and candidate items; \textbf{(2) V2} utilizes only the Active-Intent-Aware (AIA) Collaborative Path, leveraging user command-guided interaction feature extraction; \textbf{(3) V3} incorporates the complete Matcher Tool combined with the Attenuator Tool for comprehensive positive and negative preference modeling; \textbf{(4) V4} implements a simplified BGE-based semantic approach for both positive and negative preferences without personalized knowledge; and \textbf{(5) Full} represents the complete RecBot framework with the entire tool suite including the Filter Tool for hard constraint enforcement. All experiments are conducted using GPT-4.1, with results presented for Single-Round (SR) and Multi-Round (MR) scenarios. Due to space limitations, experimental results under the MRID scenario are omitted, but they yield consistent conclusions.

As illustrated in Figure~\ref{fig:offline_ablation}, Full RecBot consistently achieves superior performance across both SR and MR scenarios, particularly demonstrating efficiency in MR tasks where it attains higher pass rates while requiring fewer interaction rounds. The results reveal a clear performance hierarchy, with V1 exhibiting the weakest performance in most evaluation metrics due to its reliance solely on positive preference semantic matching, which neglects negative feedback signals that constitute the majority of user expressions in real-world scenarios and completely ignores personalized collaborative information. In contrast, V2 and V3 demonstrate substantially improved performance by incorporating user command-aware sequential modeling, with the Active-Intent-Aware collaborative component effectively bridging explicit user commands with implicit behavioral preferences.

The progression from V3 to Full RecBot illustrates the additive benefits of more effective tool integration, where the inclusion of the Filter Tool for hard constraint enforcement provides additional performance gains by ensuring strict adherence to user specifications while reducing computational overhead through candidate space pruning. These ablation results conclusively validate the necessity and effectiveness of each component within our proposed tool suite, demonstrating that optimal performance requires the coordinated deployment of all modules to better capture the multifaceted nature of user preferences in interactive contexts.

\subsection{Online Experiments}

\subsubsection{Long-term Performance}
\begin{figure}[t]
\centering
\includegraphics[width=\linewidth]{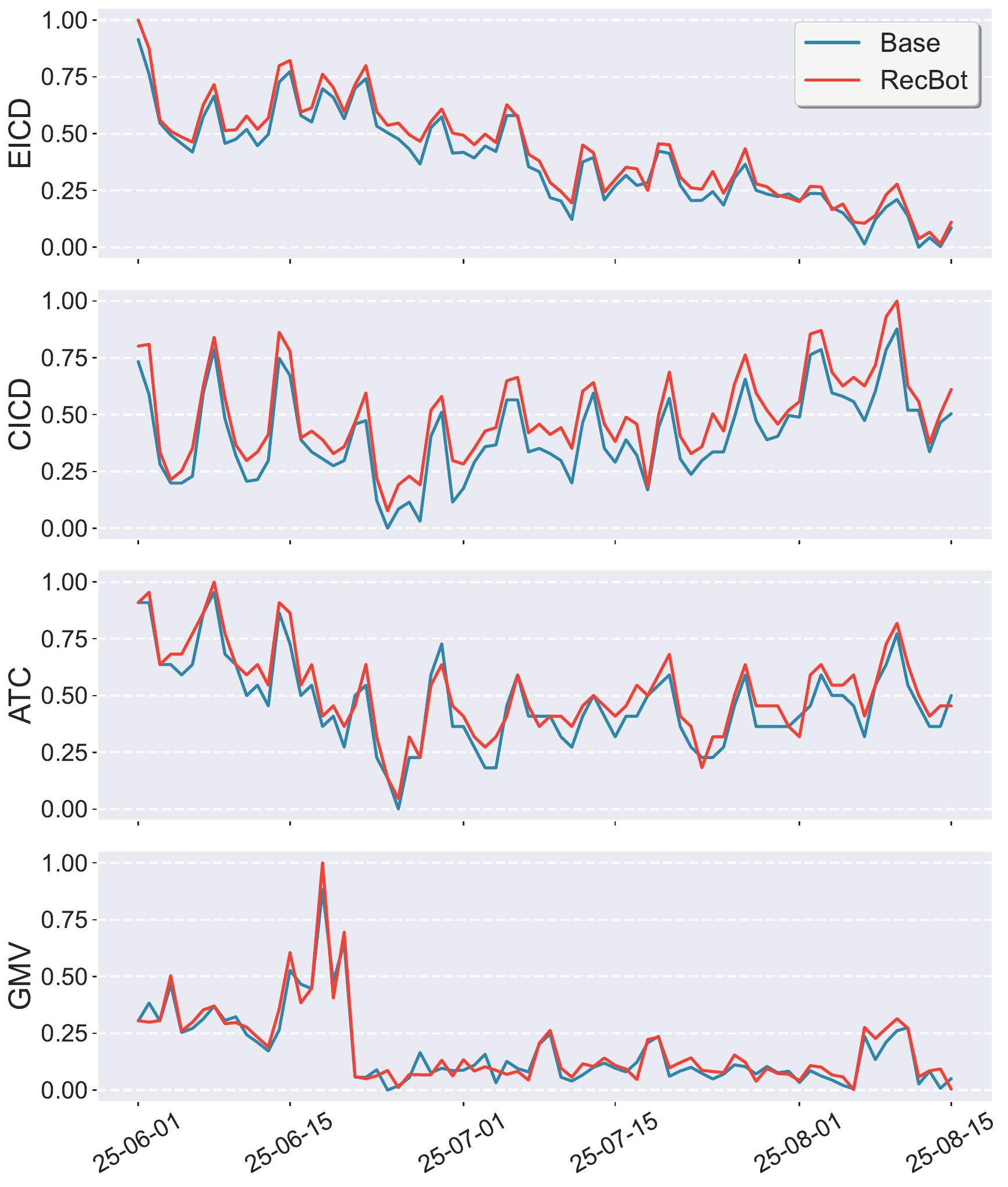}
\vspace{-2em}
\caption{Online performance curves during the three-month A/B testing period. The comparison shows RecBot \textit{vs.} the base system with all metrics normalized using min-max scaling.}
\label{fig:main_online_curve}
\end{figure}

To investigate the long-term effectiveness of the Interactive Recommendation Feed (IRF) paradigm and the RecBot framework in real-world deployment scenarios, we conducted comprehensive online A/B testing over a \textbf{three-month} period within a leading Asian e-commerce platform's homepage recommendation feed using controlled traffic allocation. Fig.~\ref{fig:main_online_curve} presents the daily performance trajectory comparison between RecBot and the baseline system, while Table~\ref{tab:online_performance} summarizes the average metric improvements across the evaluation period. To protect proprietary business information, all metric values in Fig.~\ref{fig:main_online_curve} are applied min-max normalization while preserving the relative changing trends.

The experimental results demonstrate substantial improvements across both user experience and business utility dimensions. Most notably, RecBot achieves a notable 0.71\% reduction in Negative Feedback Frequency (NFF), indicating that the framework effectively identifies user-preferred items through explicit linguistic guidance rather than relying solely on coarse-grained implicit behavioral signal interpretation characteristic of traditional recommendation paradigms. This expected improvement in user satisfaction validates the core hypothesis that enabling direct user-system communication through natural language commands substantially enhances recommendation accuracy and user experience quality.

\begin{table}[t]
\centering
\caption{Online average performance improvement of RecBot over the baseline model during the three-month A/B testing.}
\vspace{-1em}
\label{tab:online_performance}
\begin{tabular}{*{8}{c}}
\toprule
NFF$\downarrow$ & EICD$\uparrow$ & CICD$\uparrow$ & PV$\uparrow$  & ATC$\uparrow$ & GMV$\uparrow$ \\
\midrule
\textbf{-0.71\%} & \textbf{+0.88\%} & \textbf{+1.44\%} & \textbf{+0.56\%} & \textbf{+1.28\%} & \textbf{+1.40\%}   \\
\bottomrule
\end{tabular}
\end{table}

Furthermore, RecBot effectively mitigates the information cocoon effect commonly observed in homogeneous recommendation scenarios. The system demonstrates enhanced content diversity with $0.88\%$ and $1.44\%$ improvements in Exposed Item Category Diversity (EICD) and Clicked Item Category Diversity (CICD), respectively. This diversity enhancement reflects the framework's capability to capture nuanced user preferences and deliver varied content that broadens user exploration beyond historical behavioral patterns.
From a business utility perspective, RecBot delivers consistent commercial benefits with $0.56\%$, $1.28\%$, and $1.40\%$ improvements in Page Views (PV), Add-to-Cart (ATC), and Gross Merchandise Volume (GMV), respectively. These positive business outcomes demonstrate that user-centric controllable recommendations not only enhance user satisfaction but also generate tangible commercial value. The simultaneous improvement in user experience metrics and business performance validates the sustainable nature of the proposed approach.

\begin{figure}[h]
  \centering
  \includegraphics[width=\linewidth]{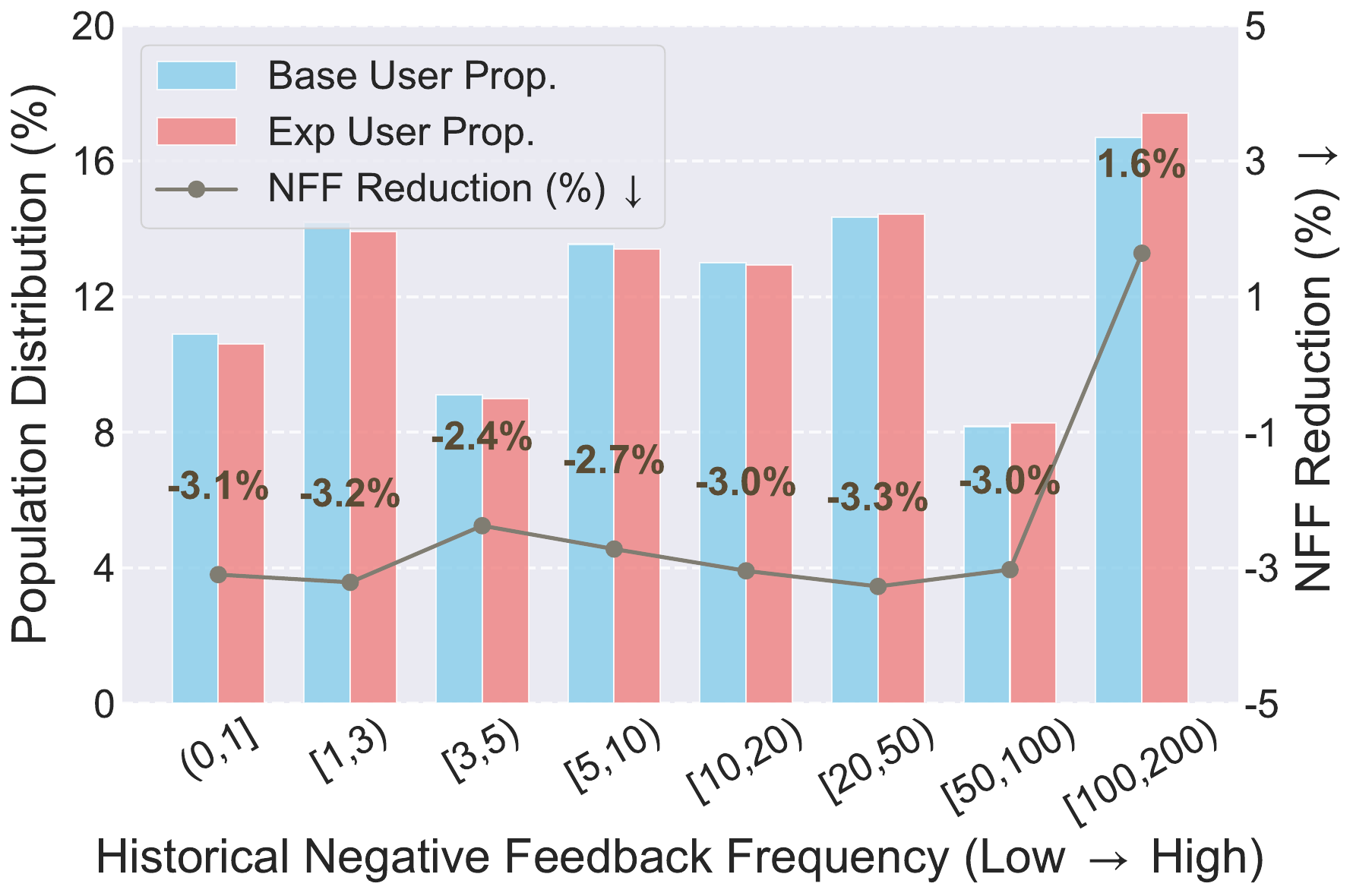}
  \vspace{-2em}
  \caption{Online performance improvements across different user groups split by historical negative feedback frequency.}
  \label{fig:online_user_group}
\end{figure}

\subsubsection{User Group Analysis}
To further investigate the differential impacts of RecBot across distinct user segments, we conducted detailed user group analysis by partitioning users based on \textit{historical negative feedback frequency}. As illustrated in Fig.~\ref{fig:online_user_group}, we established controlled experimental conditions where the baseline methods employed traditional option-based negative feedback mechanisms while our methods utilized novel language command interface. The population distributions between baseline and experimental groups remain statistically equivalent, effectively eliminating potential confounding factors arising from demographic biases.

The results demonstrate RecBot's robust effectiveness across different user groups, achieving consistent NFF reductions ranging from 2.4\% to 3.3\% for users spanning from low to moderately-high historical negative feedback frequencies. Particularly noteworthy is the [20,50) group's optimal 3.3\% improvement, indicating that the natural language manner successfully enables users with varying complaint patterns to articulate preferences more effectively than traditional implicit feedback mechanisms. The consistent improvement validates RecBot's broad applicability, though users with extremely high historical negative feedback [100,200) show a 1.6\% increase in NFF, as these chronically dissatisfied users rarely receive positive reinforcement and represent inherently challenging cases requiring more comprehensive product and algorithmic design beyond recommendation accuracy improvements.

\subsubsection{User-Command Fulfillment Analysis}
To directly assess the real-world command fulfillment capabilities of the deployed RecBot system, we conducted a detailed evaluation using both human expert annotation and an \textit{llm-as-a-Judge} approach (Qwen3-14B-Instruct fine-tuned on expert data). The evaluation focused on whether items in subsequent recommendation feeds strictly adhered to user-specified requirements following system policy adjustments, with each interaction classified as successful (complete requirement satisfaction) or failed (non-compliance). Our analysis encompassed approximately 180,000 online interaction instances. 

As presented in Table~\ref{tab:evaluation_results}, the online RecBot achieved an 88.9\% success rate according to human expert evaluation, while the automated LLM-Judge evaluation yielded a closely aligned 87.5\% success rate with 96.5\% consistency compared to human assessments. These results demonstrate that RecBot accurately interprets and responds to natural language user commands in production environments, while our LLM-based evaluation methodology provides a reliable and cost-effective alternative to manual assessment for large-scale commercial deployment monitoring.

\subsubsection{Case Study}
\begin{figure}[t]
\centering
\includegraphics[width=\linewidth]{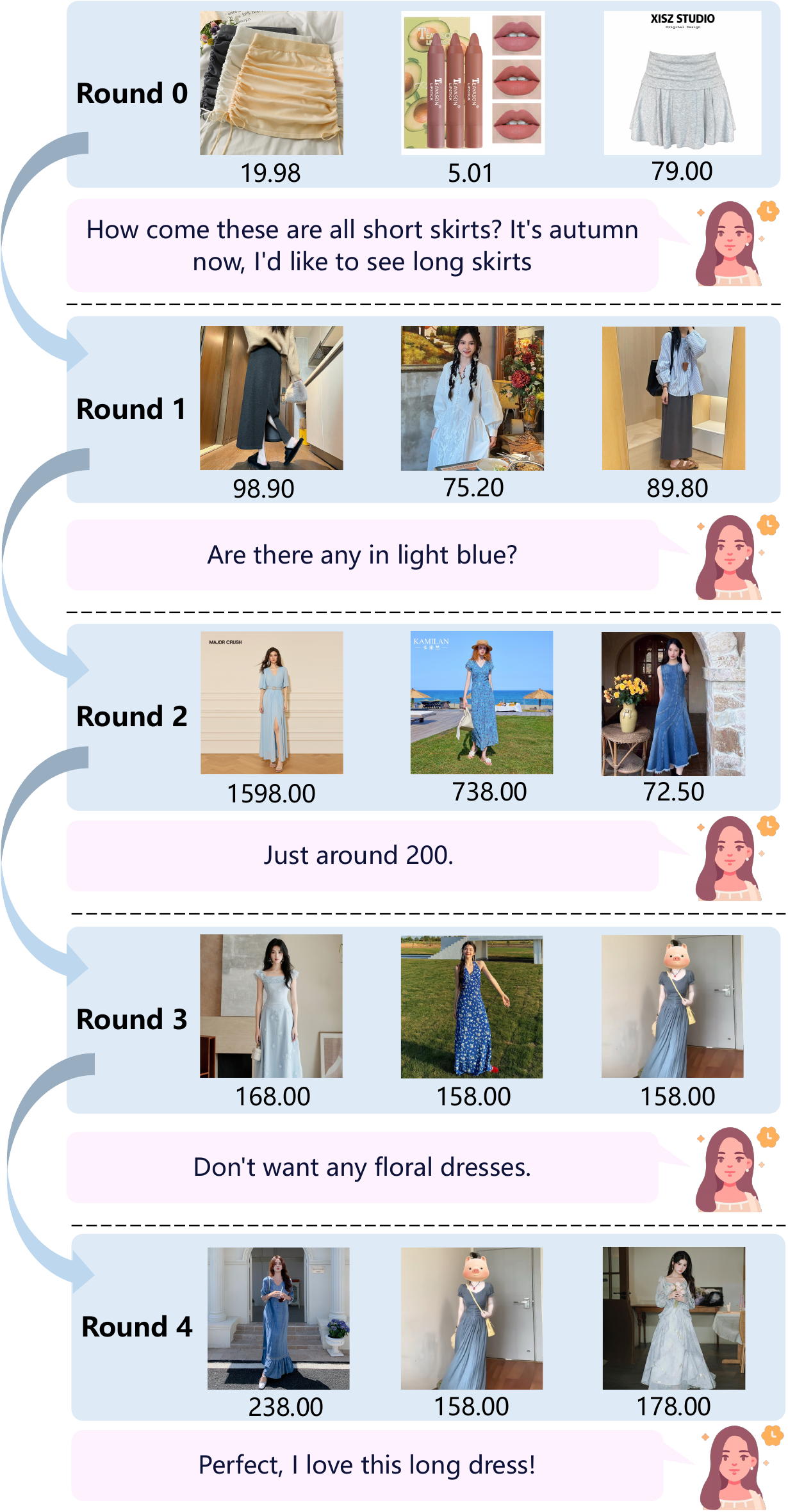}
\vspace{-1em}
\caption{Case study of RecBot on production platform.}
\label{fig:case_study}
\end{figure}

To illustrate RecBot's practical effectiveness, we present a representative multi-round interaction case from our production deployment as shown in Fig.~\ref{fig:case_study}. The user initially receives short skirts but requests ``long skirts'' for autumn weather, prompting RecBot to adapt with longer garments. Across subsequent rounds, the user progressively adds constraints: color preference (``light blue''), budget limitation (``around 200''), and negative feedback (``Don't want floral dresses''). RecBot successfully maintains all accumulated preferences while satisfying each new requirement, demonstrating effective command parsing, memory consolidation, and adaptive tool orchestration. The interaction achieves successful convergence in Round 4 with complete user satisfaction (``Perfect, I love this long dress!''), exemplifying how natural language commands enable efficient preference refinement.

\begin{table}[t]
\centering
\caption{Performance Evaluation of Online RecBot's User Command Fulfillment. The LLM-Judge accuracy is validated at 96.5\% using human evaluation as ground truth.}
\vspace{-1em}
\label{tab:evaluation_results}
\begin{tabular}{cc}
\toprule
\textbf{Evaluation Method} & \textbf{Success Rate (\%)} \\
\midrule
Human Evaluation & 88.9 \\
LLM-Judge Evaluation & 87.5 \\
\bottomrule
\end{tabular}
\vspace{-1em}
\end{table}

\section{Related Work}

\subsection{Interactive Recommendation.}
Interactive recommender systems have emerged as a critical paradigm that enables dynamic adaptation to user preferences through real-time feedback mechanisms, in contrast to traditional static approaches that rely solely on historical behavioral patterns. These systems are typically formulated as Markov Decision Processes (MDP), where recommendation policies are iteratively updated based on current system states and user responses~\cite{wu2021partially,gao2023alleviating,wu2024personalised,zhou2020interactive,chen2021large,wang2014exploration}. Recent advances have explored diverse optimization objectives and methodologies: CIRS~\cite{gao2023cirs} addresses filter bubble effects through causal reasoning-enhanced offline reinforcement learning to separate intrinsic user interests from over-exposure biases, while BiLLP~\cite{shi2024large} leverages LLMs' planning capabilities for long-term recommendation through bi-level learning mechanisms.
With the advancement of large language model agents~\cite{wang2024survey}, conversational interactive recommendation has gained prominence through natural language interfaces. InstructAgent~\cite{xu2025instructagent} introduces an LLM-mediated framework that interprets free-text user instructions and performs knowledge-enhanced re-ranking, while InteRecAgent~\cite{huang2025recommender} combines LLMs as reasoning engines with traditional recommender models as execution tools through modular toolsets including information querying, item retrieval, and ranking functions. However, these approaches are constrained to dialogue-based assistant interfaces, limiting their applicability to mainstream recommendation feed scenarios (e.g., e-commerce homepages, social media feeds) where users lack direct communication channels with the system. Our work addresses this gap by introducing the Interactive Recommendation Feed paradigm, which enables natural language commands within traditional feed-based interfaces, bridging explicit user control with scalable recommendation delivery.

\subsection{Controllable Recommender Systems.}
Controllable recommender systems have emerged as a critical research area enabling dynamic adaptation to user preferences and platform objectives without model retraining. Existing approaches primarily focus on three control dimensions: \textit{Multi-objective Control} that balances competing goals like accuracy and diversity through methods such as ComiRec~\cite{cen2020controllable} and CMR~\cite{chen2023controllable}; \textit{User Portrait Control} enabling preference specification via profile editing as demonstrated in LACE~\cite{mysore2023editable} and TEARS~\cite{penaloza2024tears}; and \textit{Scenario Adaptation Control} for cross-context variations exemplified by HyperBandit~\cite{shen2023hyperbandit} and Hamur~\cite{li2023hamur}. These methods employ diverse technical solutions including hypernetwork-based parameter generation (CMR~\cite{chen2023controllable}, HyperBandit~\cite{shen2023hyperbandit}), Pareto optimization (MoFIR~\cite{ge2022toward}), and recent LLM integration (RecLM-gen~\cite{lu2024aligning}, LangPTune~\cite{gao2024end}). Control paradigms vary by stakeholder orientation, with user-centric methods like UCRS~\cite{wang2022user} empowering individuals, while platform-mediated solutions such as CCDF~\cite{zhang2024practical} focus on algorithmic adjustments for business objectives. Unlike these approaches that need predefined control mechanisms, our method enables users to directly communicate personalized requirements through free-form natural language commands, facilitating real-time preference refinement and enhanced user control over recommendation policies.

\section{Conclusion}

In this paper, we identify fundamental limitations in traditional recommender systems that constrain user expression through passive feedback paradigms, leading to systematic misalignment between user intentions and system interpretations. To address these challenges, we introduce the Interactive Recommendation Feed (IRF) paradigm and develop RecBot, a multi-agent framework that enables user-controllable recommendation experiences through natural language commands. Comprehensive evaluation across three datasets and long-term online deployment validate the effectiveness of our proposed approach, achieving substantial improvements in both user satisfaction and business outcomes.
In the future, we will focus on developing online learning mechanisms for continuous agent evolution through online user feedback, enhancing personalized reasoning capabilities, and extending toward more intelligent interactive recommender systems with proactive anticipation and explanatory capabilities.

\clearpage


\bibliographystyle{ACM-Reference-Format}
\bibliography{sample-base}

\appendix

\end{document}